\documentclass{mn2e}
\usepackage[dvips]{graphicx}

\begin{document}

\title[A multiphase model with SN energy feedback]
{Feedback and metal enrichment in cosmological SPH simulations  -- II. A multiphase model with supernova energy feedback}

\author[Scannapieco et al.]{C. Scannapieco $^{1,2}$\thanks{E-mail: cecilia@iafe.uba.ar (CS);
patricia@iafe.uba.ar (PBT); swhite@mpa-garching.mpg.de (SDMW); volker@mpa-garching.mpg.de (VS)},
P.B. Tissera$^{1,2}$\footnotemark[1], S.D.M. White$^{3}$\footnotemark[1] and V. Springel$^{3}$\footnotemark[1] \\
$^1$ Instituto de Astronom\'{\i}a y F\'{\i}sica del Espacio, Casilla de Correos 67,
Suc. 28, 1428, Buenos Aires, Argentina.\\
$^2$  Consejo Nacional de Investigaciones Cient\'{\i}ficas
y T\'ecnicas, CONICET, Argentina.\\ 
$^3$ Max-Planck Institute for Astrophysics, Karl-Schwarzchild Str. 1, D85748, Garching, Germany.}

\maketitle

\begin{abstract}
  We have developed a new scheme to treat a multiphase interstellar medium in
  smoothed particle hydrodynamics simulations of galaxy formation. This scheme
  can represent a co-spatial mixture of cold and hot ISM components, and is
  formulated without scale-dependent parameters. It is thus particularly
  suited to studies of cosmological structure formation where galaxies with a
  wide range of masses form simultaneously.  We also present new algorithms
  for energy and heavy element injection by supernovae, and show that together
  these schemes can reproduce several important observed effects in galaxy
  evolution.  Both in collapsing systems and in quiescent galaxies our codes
  can reproduce the Kennicutt relation between the surface densities of gas
  and of star formation. Strongly metal-enhanced winds are generated in both
  cases with ratios of mass-loss to star formation which are similar to those
  observed. This leads to a self-regulated cycle for star formation activity.
  The overall impact of feedback depends on galaxy mass.  Star formation is
  suppressed at most by a factor of a few in massive galaxies, but in low-mass
  systems the effects can be much larger, giving star formation an episodic,
  bursty character. The larger the energy fraction assumed available in
  feedback, the more massive the outflows and the lower the final stellar
  masses. Winds from forming discs are collimated perpendicular to the disc
  plane, reach velocities up to $\sim 1000\,{\rm km\, s^{-1}}$, and
  efficiently transport metals out of the galaxies. The asymptotically unbound
  baryon fraction drops from $>95$ per cent to $\sim 30$ per cent from the least to the most
  massive of our idealized galaxies, but the fraction of all metals ejected
  with this component exceeds $60$ per cent regardless of mass. Such winds
  could plausibly enrich the intergalactic medium to observed levels.
   \end{abstract}

\begin{keywords}galaxies: formation - evolution - abundances  - cosmology: theory  -
methods: N-body simulations 
\end{keywords}

\section{Introduction}

Within the current cosmological paradigm, galaxies form in a hierarchical
fashion. Small systems collapse first, and larger systems build up through
accretion both of diffuse material and of pre-existing objects.  The variety
of physical processes involved in this assembly, their complex nonlinear
interplay, and the broad range of scales over which they act, make detailed
study of galaxy formation a challenging task.  Numerical simulations which can
treat both collisionless and dissipational dynamics provide a powerful tool to
address this problem. With simple recipes for star and black hole formation
and for feedback from supernovae (SNe), stellar winds and active galactic nuclei
(AGN), they allow physically
based studies of galaxy formation from the initial conditions posited by the
concordance Lambda Cold Dark Matter ($\Lambda$CDM) cosmology.  However, significant numerical challenges
arise in this approach because of the need to describe simultaneously the
large-scale dynamics of protogalactic assembly and the small-scale processes
associated with the formation and evolution of stars.

A major limitation of the popular smoothed particle hydrodynamics technique
(SPH; Gingold \& Monaghan 1977; Lucy 1977) is its inability to represent
multiphase, multiscale mixtures like the interstellar medium (ISM) in
star-forming galaxies. In the SPH formalism, the density associated with a
given particle is estimated by averaging over all neighbours within an
adaptively defined smoothing region.  In a multiphase ISM most of the volume
is filled with hot, diffuse gas while most of the mass lies in cold, dense
clouds with sizes and masses which cannot be resolved in a galaxy-scale
simulation. The density associated with 'hot' particles is then
overestimated because some 'cold' particles lie within the smoothing kernel.
This results in overestimation of the associated cooling rate, excessive
condensation of cold gas, and too high a star formation rate (Thacker et
al. 2000; Pearce et al.  2001).  Several ad hoc 'solutions' of
differing complexity have been suggested for this problem. 
Hultman \& Pharasyn (1999) used a two-phase ISM
which comprises a hot gas component and cold clouds. These two phases interact
with each other through radiative cooling and evaporation of the cold clouds.
Pearce et
al. (1999, 2001) explicitly decoupled cold and hot phases defined by pre-set
characteristic temperature boundaries. Thacker \& Couchman (2001) reduced
overcooling by allowing no radiative losses for 30 Myr from particles which
had just been heated by SNe, thereby allowing a wind to develop.
Springel \& Hernquist (2003) developed an analytic subresolution model for
the regulation of star formation in a multiphase ISM, and inserted galactic
winds of given velocity and mass-loss rate 'by hand'. 
More recently, Harfst, Theis \& Hensler (2006) also used a two-phase ISM
and assumed a variable star formation efficiency depending on the ISM properties.
In this paper, we
present a new multiphase SPH scheme, similar to that of Marri \& White (2003),
in which particles with very different thermodynamic variables do not see each
other as neighbours.  Our model decouples phases with very different specific
entropies, based on a local comparison of particle pairs.  This allows hot,
diffuse gas to coexist with cold, dense gas without introducing ad hoc
characteristic scales.

The relevance of SN feedback for the evolution of galaxies has
been emphasized by numerous observational and theoretical studies.  SNe are
the main source of heavy elements in the Universe and the presence of such
elements substantially enhances the cooling of protogalactic gas (e.g. White
\& Frenk 1991).  On the other hand, the release of energy by SNe heats up the
surrounding material, leading to disruption of cold gas clouds and to reduced
star formation. This energy can drive enriched material into the outer regions
of galaxies or even transport it into the intergalactic medium (IGM; 
Lehnert \& Heckman 1996; Dahlem, Weaver \& Heckman 1998; Rupke, Veilleux \&
Sanders 2002;  Frye, Broadhurst \& Ben\'{\i}tez 2002;
Martin 2004; Shapley et al. 2004).  Small systems are thought to be more strongly affected by SNe
because of their shallower potential wells which are less efficient in
retaining baryons (Larson 1974; White \& Rees 1978; Dekel \& Silk 1986; White
\& Frenk 1991).  The joint action of chemical enrichment and hydrodynamic
heating by SNe is often referred to as `feedback'. Feedback structures the
ISM, establishes a self-regulated cycle for star formation, and ultimately
explains the detailed luminosities of galaxies as well as the origin of heavy
elements in the IGM and elsewhere. Recent work has emphasized that feedback
from AGN may also play a critical role in shaping galaxy evolution (e.g.  Di
Matteo,Springel \& Hernquist 2005; Croton et al 2006). For
simplicity, we do not consider such processes in the current paper.

The treatment of SN feedback in simulations of galaxy formation is complicated
by the fact that the physical mechanisms which inject energy and heavy
elements into the ISM act on unresolved scales.  Consequently,
recipes must be implemented that correctly mimic their combined effects on
scales that are resolved.  In recent years, a number of authors have developed
star formation and feedback recipes for simulating galaxy formation with both
mesh-based and SPH codes (e.g. Katz \& Gunn 1991; Cen \& Ostriker 1992;
Navarro \& White 1993; Metzler \& Evrard 1994; Yepes et al. 1997;  Cen \& Ostriker 1999;
Sommer-Larsen, Gelato \& Vedel 1999; Kay et al.  2002; Lia, Portinari \&
Carraro 2002; Semelin \& Combes 2002; Marri \& White 2003; Springel \&
Hernquist 2003).  These studies have shown that simply injecting the energy
into the thermal reservoir of surrounding gas is ineffective, producing
negligible effects on the hydrodynamics.  This is because stars form (and
explode) in high-density regions where gas cooling times are short.  The
injected energy is thus radiated before it can drive significant motions (Katz
1992).  Such thermal feedback is unable to regulate the star-formation
activity or to drive the kind of winds observed in starbursting galaxies.

Several alternative schemes have been proposed to produce more effective
feedback in simulations. For example, Navarro \& White (1993) suggested
investing the SN energy directly in outward motions imposed on surrounding
gas.  Gerritsen \& Icke (1997) turned off cooling for a brief period after
particles acquire feedback energy, allowing them to evolve only adiabatically
over this time (see also Mori et al. 1997).  
Thacker \& Couchman (2000) implemented a similar recipe,
adjusting the density of heated particles in order to prevent immediate energy
losses.  More recently, Springel \& Hernquist (2003) added galactic winds to
their multiphase model by explicitly creating wind particles at a rate
proportional to the star formation rate and assigning them a predefined
outward ``wind'' velocity.  While each of these approaches has had some
success, it is clear that all have arbitrary  ad hoc elements and none is
fully satisfactory. In particular, all of them introduce characteristic values
for wind speeds, mass injection rates, flow times or other parameters which
should, in principle, be set by the dynamics of the system and which are
likely to vary substantially between systems of different scale, metallicity,
etc. Given the central importance of feedback, it remains an important task to
improve the numerical treatment of this process in our simulations.

In this paper, we implement energy and chemical feedback in the context of our
multiphase model for the ISM.  Our new energy feedback scheme supplements the
treatment of chemical enrichment already discussed in Scannapieco et
al. (2005, Paper I), and is also implemented in the parallel TreeSPH code
{\small GADGET-2} (Springel 2005). Our primary goal is to address some of the
shortcomings of previous implementations within a comprehensive and consistent
description of energy and chemical feedback which should then produce
realistic galactic outflows.  An important advantage of the new scheme is that
it avoids  ad hoc scale-dependent parameters of the kind present in most
previous work. This makes our scheme well suited to cosmological structure
formation simulations where galaxies with a wide range of properties form
simultaneously.

Our paper is organized as follows.  In Section~\ref{code}, we present the
numerical implementations of our multiphase approach and of the energy
feedback model, testing their performance using idealized simulations of disc
galaxy formation.  In Section~\ref{results}, we use similar collapse
simulations to discuss the impact of SN feedback on galaxy formation. We
include an analysis of the effects on star formation and the generation of
winds (Section~\ref{sfr-outflows}), on chemical enrichment of the ISM and the
IGM (Section~\ref{metals}), and within galaxies of differing total mass
(Section~\ref{mvirial}).  In Section~\ref{kennicutt}, we analyse both such
collapse simulations and simulations of equilibrium galaxies similar to the
Milky Way, exploring whether a suitable choice of star formation efficiency
allows us to reproduce the observed Kennicutt relation between the surface
densities of gas and of star formation in galaxies.  Finally,
Section~\ref{conclusions} summarizes our conclusions.

\section{Numerical implementation}
\label{code}

In this section, we discuss how we treat a multiphase ISM, and implement
energy feedback from SN explosions.  We also test the performance of these new
numerical schemes, which are grafted onto the version of {\small GADGET-2}
discussed in Paper I, which includes metal enrichment by Type II SNe (SNII) and
Type Ia  SNe (SNIa) as well as metal-dependent cooling.

\subsection{Multiphase gas model}
\label{decoupling}

\subsubsection{Model description}

The ISM is known to have a complex, multiphase structure in
which interpenetrating components of similar pressure but very different
temperature and density span a wide range of spatial scales (e.g.  McKee \&
Ostriker 1977; Efstathiou 2000). This multiphase character is difficult to
represent appropriately in numerical simulations.  In particular, standard
implementations of SPH reproduce it poorly, because they have insufficient
resolution to represent clouds of cool, dense gas embedded in a hotter diffuse
medium.  Numerical artefacts result from the SPH estimate of gas density which,
in a commonly used implementation, can be written as
\begin{equation}
\label{density}
\left<\rho_i\right> =  \sum_j^N m_j W_{ij} \,, 
\end{equation}
where $j$ runs through the $N$ neighbours of particle $i$, $m_j$ denotes the
mass of particle $j$, and $W_{ij}$ is a symmetrized kernel function. Diffuse
phase particles close to a dense cloud overestimate their densities by
including cloud particles in the sum over neighbours. This leads to an
underestimation of their cooling times and so to excessive accretion of gas
on to the cloud (Pearce et al.  1999). This then artificially boosts the star
formation rate. A second, more serious problem, also caused by the lack of
resolution, is that SN feedback is not channelled into the hot phase but rather
is dumped in the dense gas surrounding star-forming regions, resulting in the
immediate radiative loss of the energy and the confinement of the metals. This
prevents the launching of galactic winds, the ejection of heavy elements, and
the self-regulation of star formation activity.

In order to address these issues and to improve the treatment of the ISM in
SPH, we have developed a multiphase scheme which allows a larger overlap of
the diffuse and dense gaseous components and can be combined with a more
effective scheme for injecting the energy and heavy elements from SNe.  The
innovation in our scheme relates to the selection of neighbours.  If two gas
particles have dissimilar thermodynamic properties, they are explicitly
prevented from being neighbours in the SPH calculations.  In detail, this
works as follows.  We {\it decouple} a given particle $j$ from particle $i$,
meaning that $j$ is explicitly excluded from the neighbour list of $i$, if the
following conditions are fulfilled:
\begin{equation}
\label{entropy_condition}
A_i > \alpha A_j\,,
\label{entropy_condition} 
\end{equation}
\begin{equation}
\label{shock_condition}
\mu_{ij} < c_{ij}\, ,
\end{equation}
where $\alpha$ is a constant, $A_i$ is the entropic function of particle $i$,
and $\mu_{ij}$\footnote{We use the artificial viscosity term as parametrized in equation~(14) of
Springel (2005).}
 and $c_{ij}$ measure the pair-averaged local velocity
divergence and sound speed, respectively.  The entropic function characterizes
the specific entropy $s$ of gas particles through
\begin{equation}
P = A(s)\rho^\gamma ,
\end{equation}
where $P$ denotes pressure and $\gamma=5/3$ is the adiabatic index for a
monoatomic ideal gas.  The second condition of equation~(\ref{shock_condition}) is
included to avoid decoupling in shock waves, which, as noted by Marri \& White
(2003), can lead to unphysical effects when particles on opposite sides of a
shock do not ``see'' each other.

The code assumes a fixed mass within the smoothing
length which corresponds to $\sim 32$ neighbours (Springel
\& Hernquist 2003).
In the context of our decoupling scheme, 
if the physical state of a given particle changes,
a new list of neighbours is calculated to assure that
the required $\sim 32$ neighbours are enclosed.

Conservation of energy and momentum requires symmetric force evaluations
between all particle pairs to ensure the validity of Newton's third
law. Hence, for the force calculation only we include the extra requirement
that if particle $i$ excludes particle $j$ from its neighbour list, then
particle $j$ does not consider particle $i$ as a neighbour either.  In this
way the symmetry of force calculations is preserved in our code.

An important advantage of the above scheme is that the decoupling criterion is
checked on a pair-wise basis. This allows a formulation which is independent of
scale or resolution-dependent parameters. Only the dimensionless free
parameter $\alpha$ needs to be set.  A multiphase structure naturally appears
in the gas component in this model once realistic cooling processes are
included. This substantially improves the representation of the
ISM in star-forming galaxies.

\subsubsection{Tests of the model}

In order to test the performance of this decoupling scheme, we carried out a
set of idealized simulations of the formation of isolated disc galaxies
similar to those in the early work of Navarro \& White (1993).  The initial
conditions are generated by radially perturbing a spherical grid of superposed
dark matter and gas particles to produce a cloud with density profile
$\rho(r)\sim r^{-1}$ and radius $100\,h^{-1}{\rm kpc}$ ($h=0.7$).  This sphere is
initially in solid body rotation with an angular momentum characterized by
spin parameter $\lambda \simeq 0.1$. The initial thermal energy of the system
is only $5$ per cent of its binding energy, i.e.~the gas is cold.  We have simulated a
$10^{12}\,h^{-1}{\rm M}_\odot $ mass system, 10 per cent of which is in the
form of baryons, and used $\sim 9000$ particles for the gas and $\sim 9000$ for the
dark matter.  This yields particle masses of $\sim 10^8\,h^{-1}{\rm M}_\odot$
for dark matter and $10^7\,h^{-1}{\rm M}_\odot$ for gas.  We adopted
gravitational softening lengths of $1.50$  and $0.75\ h^{-1}$ kpc
for dark matter and gas particles, respectively. In the tests
presented in this section, we do not allow star formation, and the cooling 
assumed primordial element abundances.

We ran two simulations of the system described above, the only difference
being the inclusion of our multiphase treatment. In the simulation
including the multiphase model, we have adopted a decoupling parameter of
$\alpha = 50$. 
In order to better highlight
the effects of the multiphase model, we specify two reference phases, defined
by $A \geq A_{\rm crit}$ for the 'hot' component, and $A < A_{\rm crit}$ for
the 'cold' component, where $A_{\rm crit}$ is the entropic function corresponding to a
temperature of $T=2\,T_*$ and a density of $\rho = 0.1\ \rho_*$.  
Here $T_* = 4 \times 10^4$ K  and $\rho_* = 7 \times 10^{-26}$ g cm$^{-3}$ 
are the star formation
thresholds assumed in Paper I. We stress that these definitions are not
related to the decoupling model and are only introduced to facilitate
presentation of our results.

\begin{figure}
\includegraphics[width=90mm]{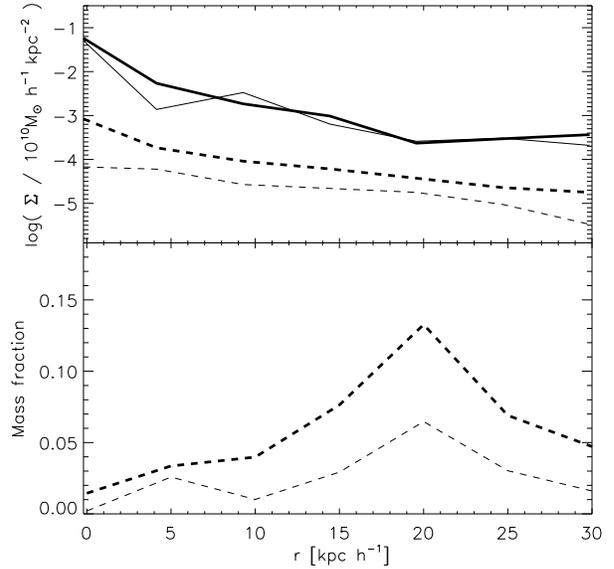}
\caption{Top: Surface density profiles projected on to the disc plane after 
$0.9\, h^{-1}{\rm Gyr}$ for hot (dashed
  lines) and cold (solid lines) gas in our cooling-only simulations of
  isolated collapsing spheres carried out using a standard SPH implementation
  (thin lines) and using our decoupling scheme  with $\alpha=50$ (thick lines). Bottom: The
  fraction of baryonic mass belonging to the hot medium as a function of radius for the
  same two simulations. These plots are restricted
to the disc plane with a vertical distance of  $|z|<2\, h^{-1}{\rm kpc}$.}
\label{density-no-SF}
\end{figure}

In the upper panel of Fig.~\ref{density-no-SF} we compare the surface gas density
profiles, projected on to the disc plane, of the two simulations after 
$0.9\,h^{-1}{\rm Gyr}$. Profiles for the
'hot' and 'cold' phases defined above are shown separately.  In the lower
panel we plot the fraction of baryonic mass which is ``hot'' as a function of
radius.
These plots have been made taking into account the gas within 
a vertical distance  $|z|<2$ $h^{-1}$ kpc in order to restrict to the
disc plane.
The surface density profiles for the cold gas (solid lines) are almost
indistinguishable, but the amount of hot gas (dashed lines) is greater at all
radii when phase decoupling is included.  From the lower panel we can see that
the hot medium has grown at the expense of the cold one. This is a result of
the elimination by the decoupling algorithm of the excess cooling discussed
above\footnote{The entropy-conserving formulation of SPH (Springel \&
Hernquist 2002) implemented in {\small GADGET-2} has solved the 
extreme problem of a near-absence of hot particles in the central regions 
found by Marri \& White~(2003) when carrying out a similar test using
codes based on {\small GADGET-1}.}.

We have investigated the sensitivity of the decoupling scheme to the value of
the free parameter $\alpha$ in equation~(\ref{entropy_condition}).  Too small a
value for $\alpha$ would lead to decoupling of gas particles with very similar
properties, while too large a value would produce no effect at all.  We ran
tests varying the value of $\alpha$ over the range $\alpha=5$ to $\alpha=100$,
and found that our results are insensitive to the actual value within this
range. This reflects  the large entropy difference between the various
phases of a realistic ISM caused by the structure of the radiative cooling
function. We will generally choose $\alpha=50$ for later experiments in this
paper.

\vspace{0.5cm}

\subsection{Energy feedback by supernovae}
\label{feedback}

\subsubsection{ Model description}

Our model for energy feedback by SNe resorts to an explicit segregation of the
gas surrounding a star particle into a cold dense phase and a diffuse phase,
both for the release of SN energy and for the mixing of newly synthesized
elements into the ISM.  The separation into two phases, and the different
treatment of them with respect to feedback, is motivated by previous attempts
to model the release of SN energy, which have repeatedly found that this is
ineffective if direct thermalization is invoked because the cold dense gas in
which stars form has a very short cooling time (Katz 1992). In addition, the
characteristic physical scales relevant to feedback in the real ISM are
usually not resolved in galaxy formation simulations (Thacker \& Couchman
2001).

Surrounding each star particle with exploding SNe we define two gaseous phases
which we loosely denote {\it cold} and {\it hot}.  The {\it cold} phase
consists of gas with $T < 2 \,T_*$ and $\rho > 0.1\rho_*$, while the rest of
the gas is considered to be part of the {\it hot} phase, even though much of
it may actually have $T<2\, T_*$.  These two phases are treated differently at
the time of metal and energy distribution.  The values of $ \rho_*$ and $T_*$
($7 \times 10^{-26}$ g cm$^{-3}$ and $4 \times 10^4$ K, respectively) are
typical of local star-forming regions and are assumed to be independent of the
global properties of the systems.  Note that these phases are not directly
related to the decoupling scheme itself, or to the hot and cold gas
components we used in the previous section to analyse some of our results.

In our approach,  stars are assigned  two
different smoothing lengths, corresponding to the  cold and hot phases. 
Both smoothing lengths  are adjusted to
enclose the same mass (we have assumed the mass corresponding to $\sim 32$
neighbours).  
In the case of an extremely diffuse, hot phase, an unphysically large smoothing
length could be required to enclose $\sim 32$
neighbours. 
In order to avoid this behaviour, we  set for each star particle a
maximum length to search for hot neighbours, which
we assumed to be $10$ times the smoothing length calculated for its
cold neighbours.
We have tested the sensitivity of our results to this choice
running simulations assuming
different maximum lengths allowed ($\sim 10-100$ times the cold smoothing length), finding
that they are not affected. We stress the fact that
the neighbour searches for stars have no relation to the SPH treatment
for the gas.

Conceptually, we can divide our discussion of the SN energy feedback model
into two main stages: energy production and energy release. We shall discuss
them in turn.

To describe energy production, we have included the energy contributions
associated with SNII and SNIa explosions.  Briefly, our scheme works as
follows: at each integration time-step, we calculate the number of SNII by
adopting an initial mass function (IMF) and by assuming that stars more
massive than $8$ M$_{\odot}$ end their lives as SNII after $\approx 10^6\,{\rm
yr}$.  For SNIa, we use the W7 model of Thielemann, Nomoto \& Hashimoto
(1993).  In order to estimate the SNIa number, we adopt an observationally
motivated relative rate with respect to SNII, and we assume a progenitor
lifetime in the range $[0.1,1]\,{\rm Gyr}$ as described in Paper I.  Finally,
we assume that each SN injects $10^{51}$ ergs of energy into the surrounding
gas.

In order to distribute the energy produced by a given star particle, we
separately identify the gaseous neighbours that belong to the {\it hot} and to
the {\it cold} phases.  Smoothing lengths are calculated for the star particle
so that the number of neighbours in each of these phases is similar to the
number used in the SPH formalism to estimate hydrodynamical forces.  Within
each phase, gaseous neighbours then receive a fraction of the SN energy,
weighted by the appropriate smoothing kernel.

We assume that a fraction $\epsilon_{\rm h}$ of the SN energy is instantaneously
thermalized and appears in the hot phase of the ISM around the star particle,
while a fraction $\epsilon_{\rm r}$ is directly radiated away by the cold phase.
The remaining fraction of the energy, $\epsilon_{\rm c} = 1 - \epsilon_{\rm h} -
\epsilon_{\rm r}$, is injected into the cold gas causing some of it to join the hot
phase. Because there is no consensus on the phenomenology of the ISM which
could provide specific values for these fractions, we assume them to be
independent of local conditions and we treat them as free parameters to be
adjusted so as to reproduce the observations of star-forming systems.  We vary
their values widely in order to analyse their impact on the dynamics of our
model.  Choosing $\epsilon_{\rm r} = 1$ implies that all the SN energy is radiated
away by the cold phase with no impact on the overall dynamics.  Examples of
the evolution in this extreme case are given by our multiphase-only runs (e.g. 
D12 in Table 1).  For the sake of simplicity, we have chosen
$\epsilon_{\rm r} = 0$ in the rest of our tests, noting that a non-zero value could
always be compensated by a different IMF or different assumed SN energies. In
this case, the effects of feedback depend only on $\epsilon_{\rm h}$ (or $\epsilon_{\rm c}$). In practice,
both $\epsilon_{\rm h}$ and $\epsilon_{\rm r}$ should be varied in order to get the best
possible match to observation.

For each gas particle belonging to the cold phase, we define a {\it reservoir}
in which the SN energy that the particle receives from successive SN
explosions is accumulated.  Once the accumulated energy ($E_{\rm res}$) is
high enough to modify the thermodynamic properties of a cold gas particle in
such a way that its new properties will resemble those of the hot phase, we
{\it promote} the cold particle, dumping its reservoir energy into its
internal energy.  This {\it promotion scheme} has been developed in order to
prevent artificial losses of the SN energy by the cold phase and to ensure
that cold ISM gas can be entrained in outflows, producing {\it mass-loaded}
galactic winds.

In order to decide whether the promotion of a given cold particle should take
place, we consider the following physical conditions: (i) the final physical
state we want the particle to reach and (i) the energy needed for the process
to take place, given its current physical state.  Since we want to mimic the
transformation of cold gas into gas that is in thermal equilibrium with the
hot-gas surroundings, we require that the final state for a cold particle that
is promoted should be similar to that of its local hot phase environment.
Based on our decoupling scheme, the local hot phase of any given cold gas
particle is naturally defined by those neighbours from which the cold particle
is decoupled (see Section~\ref{decoupling}).

Assuming that the transformation from the cold to the hot phase takes place at
constant pressure, we define a {\it promotion energy} for each particle.  This
promotion energy ($E_{\rm prom}$) takes into account the energy needed not
only to raise the temperature of the particle to that of its hot neighbours,
but also for it to expand against the ambient pressure to reach their mean
density\footnote{The internal energy of a particle $i$  is described as 
$U_i = (\gamma-1)^{-1}\   m_i 
A_i  \rho_i^{\gamma-1}$. Since we consider a process at constant pressure, we need
  energy  not only to raise the  temperature of the particle but also to
work against the background pressure as it expands.
 Expressed differently: ${\rm d}Q = {\rm d}U +
P {\rm d}V$ with $P{\rm d}V = (\gamma-1) {\rm d}U$, which results in  equation~\ref{Eprom}.}.  
Thus, we require the promoted particle to have an entropic function at
least as high as the mean of those of its hot neighbours ($A_{\rm Avg}^{\rm
hot}$). The conditions for promotion are the following:
\begin{equation}
E_{\rm res} > E_{\rm prom} = {\gamma \over {\gamma-1}}\  m_i \Big[ A_{\rm Avg}^{\rm hot}\ (\rho_{\rm Avg}^{\rm hot})^{\gamma -1} -  A_i\ \rho_i^{\gamma -1} \Big]
\label{Eprom}
\end{equation}
\begin{equation}
A_{\rm new} > A_{\rm Avg}^{\rm hot}  
\label{prom2}
\end{equation}
Here, $A_{\rm new}$ is an estimate of the new entropic function of the
particle after promotion and $A_{\rm Avg}^{\rm hot}$ and $\rho_{\rm Avg}^{\rm
hot}$ are the mean entropic function and mean density of the hot neighbours
surrounding the cold gas particle, respectively.
The value of  $A_{\rm new}$ is calculated assuming that the  energy 
of the particle after promotion will be its actual energy plus the reservoir 
and that the new density  will be the average density of the
hot neighbours, $A_{\rm Avg}^{\rm hot}$.

This scheme guarantees that once a gas particle is promoted, it will have
thermodynamic properties matching those of its local hot environment.
Consequently, it will remain hot at least as long as nearby hot material.  The
procedure also ensures that the smooth character of SPH is preserved and no
`flip-flop' instabilities are generated during promotion.  Note that the
promotion criterion is evaluated for each cold gas particle individually, and
that it depends both on the particle's own thermodynamic properties and on
those of its local hot environment.  This scheme allows us to mimic the
reheating of cold interstellar gas by SN explosions, providing a link between
SN energy feedback and our multiphase treatment, and resulting in a
self-regulated star-formation cycle.

\begin{figure*}
\includegraphics[width=120mm]{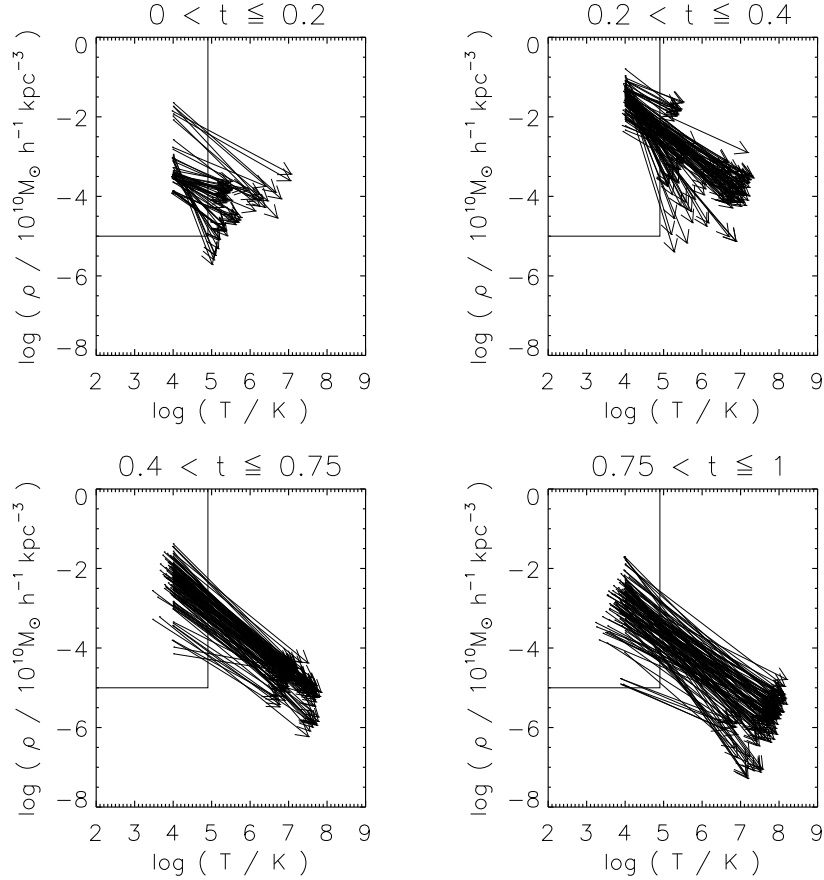}
\caption{Temperature-density distribution of promoted particles in 
  F12-0.5, which corresponds to a $10^{12}\, h^{-1} {\rm M}_\odot$ mass
  system, run with energy feedback and $\epsilon_{\rm c}=0.5$.  The arrows point
  from the initial state before promotion to the final one after promotion.
  The solid lines separate the cold and hot phases of the feedback scheme.
  The different panels correspond to particles promoted in different time
  intervals, indicated in units of $h^{-1}{\rm Gyr}$.}
\label{promoted}
\end{figure*}

We also include an extra requirement for promotion in order to treat cases
where cold particles have few hot neighbours within the maximum smoothing
length allowed, for example because all gas is cold.  If this happens, we
prefer to keep the particle in the cold phase until a well-defined hot
environment has formed.  
This ensures numerical stability in cases where the
hot phase does not exist or is very poorly sampled.  
We consider the hot environment to be well defined if it is  numerically
resolved. We therefore require a minimum of hot neighbours to be identified before
a cold particle to be eligible for promotion. 
We have run simulations with different
choices and find that at least five hot neighbours are needed to
avoid instabilities and to get converged results. 
In the  experiments presented in this paper, we adopt this minimum
number.

In the case where a cold gas
particle satisfies the conditions to be transformed into a star particle when
still having a reservoir of energy, we distribute this energy to its
cold-phase neighbours using the smoothing kernel. In the extreme case
that there are no cold neighbours, the energy is assumed to be lost and the reservoir
is reset to zero. Finally, a numerical artefact could
also develop in our scheme when cold particles become 'hot' by numerical
noise (near the density and temperature thresholds segregating phases).  
In this case, the SN energy is immediately thermalized
and the reservoir is reset to zero.
If this happens,
feedback energy is often lost because these particles immediately radiate
the newly thermalised energy.  We have checked, however,  that averaged
over a simulation's duration the total reservoir
energy lost in this way is a negligible fraction ($\sim 10^{-5}$) of the
total energy produced by SN.

Finally, we have also slightly modified the chemical enrichment model of Paper
I to work in a consistent fashion with our new treatment of the SN energy. To
this end, we distribute the chemical elements into the hot and cold phases of
a given star particle  similarly to the energy distribution. 
For metal distribution we assume different 
fractions $\epsilon_{\rm c}^m$ and $\epsilon_{\rm h}^m$ so that
$\epsilon_{\rm c}^m + \epsilon_{\rm h}^m = 1$. These fractions could, in principle, be tuned differently
than the fractions for energy distribution in order to match observations.
Our model assumes that  the chemical distribution always occurs simultaneously
with SN explosions, i.e.~there is no  metal reservoir corresponding to the promotion
scheme described above. Clearly, it would be possible to choose different
fractions for distributing heavy elements than for distributing energy but, at this
stage, we prefer to avoid additional parameters  by assuming the  metal fractions
to be equal to the energy fractions in all experiments analysed in this paper.

\subsubsection{ Tests of the model}

As a first test of the dynamics of our numerical scheme, we have run a set of
simulations of the evolution of the idealized initial conditions described in
Section~\ref{decoupling}.  These experiments include metal-dependent radiative
cooling, star formation, chemical enrichment and energy feedback.  They 
have all been run with the same star formation and chemical input parameters: a
star formation efficiency $c=0.1$, a Salpeter IMF  with lower and upper
mass cut-offs of 0.1 and 40 M$_{\sun}$, respectively, a SNIa rate of $0.3$ relative
to SNII which is a typical value coming from observations in Sa-Sb galaxies 
(as used in Paper I), and a lifetime interval of [$0.1,1$] Gyr for SNIa (see Paper
I for details).  We have assumed an instantaneous recycling approximation for SNII.
We have used metal-dependent cooling functions adapted from
Sutherland \& Dopita (1993) as described in Paper I, and assume no cooling for
gas particles with temperature lower than $10^4$ K. 
The decoupling scheme
has been turned on for all the experiments with a decoupling parameter of
$\alpha=50$. 
 We have used gravitational softening lengths 
of $1.50$, $0.75$ and $1.13\,h^{-1}{\rm kpc}$  for dark matter, gas 
and star particles, respectively.

\begin{figure*}
\includegraphics[width=180mm]{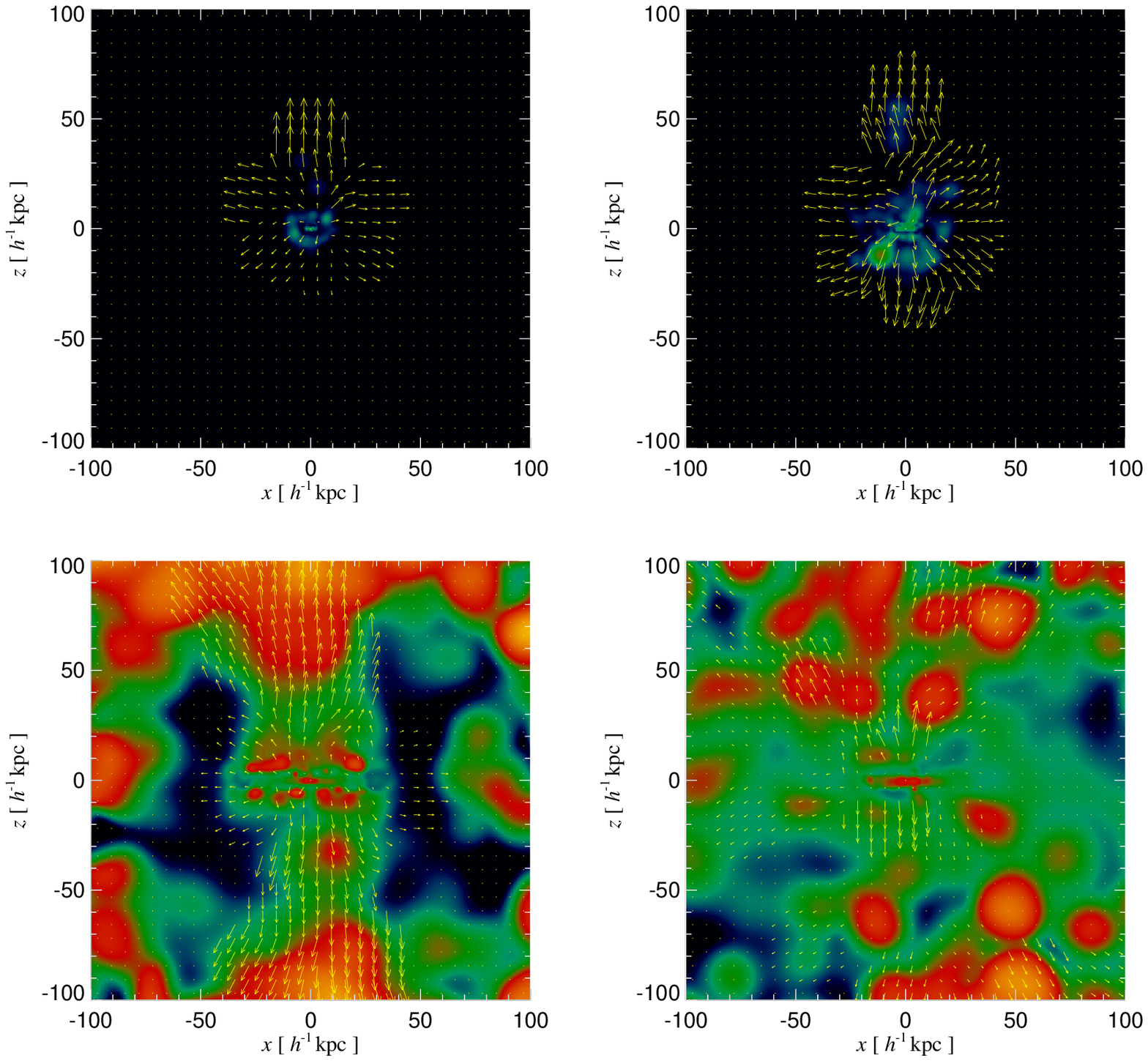}
\includegraphics[width=90mm]{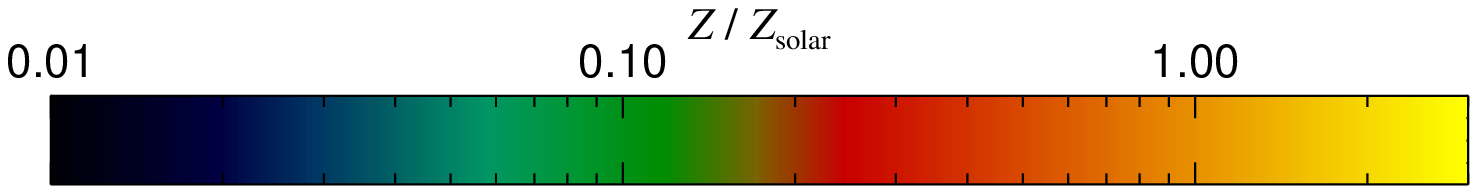}
\caption{Projected velocity field of recently promoted particles (i.e. 
particles promoted in the last $0.2\ h^{-1}$ Gyr, arrows) for
  our simulation F12-0.5 with supernova feedback with
  $\epsilon_{\rm c}=\epsilon_{\rm h}=0.5$.  The different panels correspond to different
  times in the evolution of the system: $t=0.3$ (upper left-hand panel), $0.4$
  (upper right-hand panel), $0.75$ (lower left-hand panel) and $1.0\, h^{-1}{\rm Gyr}$
(lower right-hand panel).
  The lengths of the arrows scale with the magnitude of the local velocity,
  with the longest arrows corresponding to $\sim 1000\,{\rm km\,s^{-1}}$.  The
  colour map in the background encodes the projected metallicity field,
  according to the colour-scale shown at the bottom. }
\label{outflows}
\end{figure*}

We here analyse simulations of systems of different mass, run with feedback
parameter $\epsilon_{\rm c}=0.5$.  For comparison, we have computed corresponding
reference simulations without SN feedback in all cases.  Three higher
resolution simulations have also been done to test numerical convergence.  For
the different total masses, the softening lengths as well as the initial radii
have been scaled in proportion to $M^{1/3}$, relative to the parameters
assumed for the $10^{12}\,h^{-1}{\rm M}_\odot$ system, leading to identical
collapse times and characteristic densities for all systems.  The main
properties of these simulations are summarized in Table~\ref{simus-feedback}.

\begin{table*}
\caption{Main properties of our simulations of idealized protogalaxy
  collapses. The columns give total mass, initial number of gas and
  dark matter particles, and the feedback input parameter ($\epsilon_{\rm c}$). We
  also list the final stellar mass fraction, the fraction of unbound baryons,
  the fraction of metals locked into the hot gas, and the fraction of metals
  locked into stars at $t = 1.0\, h^{-1}{\rm Gyr}$.}
\begin{center}
\begin{tabular}{lccccccc }\hline
Test &  $M_{\rm vir}$  $(h^{-1}{\rm M}_\odot)$ &  $N_{\rm gas}=N_{\rm dark}$ & $\epsilon_{\rm c}$ &
$M_{\rm star}/M_{\rm bar}$ & $M_{\rm unbound}/M_{\rm bar}$& $f_{\rm met}^{\rm hot}$
& $f_{\rm met}^{\rm stars}$\\\hline

D12        &   $10^{12}$  &   9000   &    -       & 0.581 & 0.025 & 0.002 & 0.557\\
F12-0.1    &   $10^{12}$    &   9000 &   0.1      & 0.324 & 0.411 & 0.910 & 0.045 \\
F12-0.5    &   $10^{12}$    &   9000 &   0.5      & 0.362 & 0.309 & 0.556 & 0.212\\
F12-0.9    &   $10^{12}$    &   9000 &   0.9      & 0.231 & 0.315 & 0.710 & 0.188\\ \\

D11.5        &   $10^{11.5}$  &   9000   &    -   & 0.738 & 0.002 & 0.000 & 0.663\\
F11.5-0.1  &   $10^{11.5}$  &   9000 &  0.1       & 0.415 & 0.361 & 0.903 & 0.050\\
F11.5-0.5  &   $10^{11.5}$  &   9000 &  0.5       & 0.390 & 0.441 & 0.755 & 0.170\\
F11.5-0.9  &   $10^{11.5}$  &   9000 &  0.9       & 0.276 & 0.613 & 0.793 & 0.152\\\\

D11    &   $10^{11}$    &   9000 &  -             & 0.824 & 0.000 & 0.000 & 0.703\\
F11-0.1  &   $10^{11}$  &   9000 &  0.1           & 0.398 & 0.364 & 0.904 & 0.044\\
F11-0.5    &   $10^{11}$    &   9000 &  0.5       & 0.403 & 0.407 & 0.694 & 0.169\\
F11-0.9  &   $10^{11}$  &   9000 &  0.9           & 0.273 & 0.666 & 0.813 & 0.146\\\\

D10.5  &   $10^{10.5}$  &   9000 &  -             & 0.712 & 0.000 & 0.000  & 0.608\\
F10.5-0.1  &   $10^{10.5}$  &   9000 &  0.1       & 0.342 & 0.416 & 0.903 & 0.043\\
F10.5-0.5  &   $10^{10.5}$  &   9000 &  0.5       & 0.318 & 0.525 & 0.780 & 0.141\\
F10.5-0.9  &   $10^{10.5}$  &   9000 &  0.9       & 0.207 & 0.779 & 0.881 & 0.119\\\\

D10        &   $10^{10}$    &   9000 &  -         & 0.520 & 0.000 &  0.000 & 0.426 \\
F10-0.1    &   $10^{10}$    &   9000 &  0.1       & 0.271 & 0.501 &  0.911 & 0.031\\
F10-0.5    &   $10^{10}$    &   9000 &  0.5       & 0.258 & 0.633 &  0.848 & 0.107\\
F10-0.9    &   $10^{10}$    &   9000 &  0.9       & 0.174 & 0.819 &  0.901 & 0.085\\\\

D9.5      &   $10^{9.5}$  &   9000 &  -           & 0.451 &  0.000 & 0.000 & 0.405 \\
F9.5-0.1  &   $10^{9.5}$  &   9000 &  0.1         & 0.155 &  0.548 & 0.913 & 0.017 \\
F9.5-0.5  &   $10^{9.5}$  &   9000 &  0.5         & 0.184 &  0.821 & 0.899 & 0.101 \\
F9.5-0.1  &   $10^{9.5}$  &   9000 &  0.9         & 0.113 &  0.902 & 0.945 & 0.051 \\\\

D9     &   $10^{9}$    &   9000 &  -              & 0.373 & 0.004 & 0.000 & 0.382\\
F9-0.1  &   $10^{9}$  &   9000 &  0.1             & 0.107 & 0.596 & 0.912 & 0.011\\
F9-0.5    &   $10^{9}$    &   9000 &  0.5         & 0.041 & 0.662 & 0.840 & 0.031\\
F9-0.9  &   $10^{9}$  &   9000 &  0.9             & 0.069 & 0.927 & 0.950 & 0.044\\\\

F12-0.5-H    &   $10^{12}$    &   40000 &   0.5   & 0.365 & 0.386 & 0.735 & 0.204\\
F10-0.5-H    &   $10^{10}$  &   40000  &   0.5    & 0.240 & 0.326 &  0.782 & 0.095 \\
F9-0.5-H    &   $10^{9}$   &   40000  &   0.5     & 0.061 & 0.924 & 0.954 & 0.036\\
\hline
\end{tabular}
\label{simus-feedback}
\end{center}
\end{table*}

\begin{figure*}
\includegraphics[width=180mm]{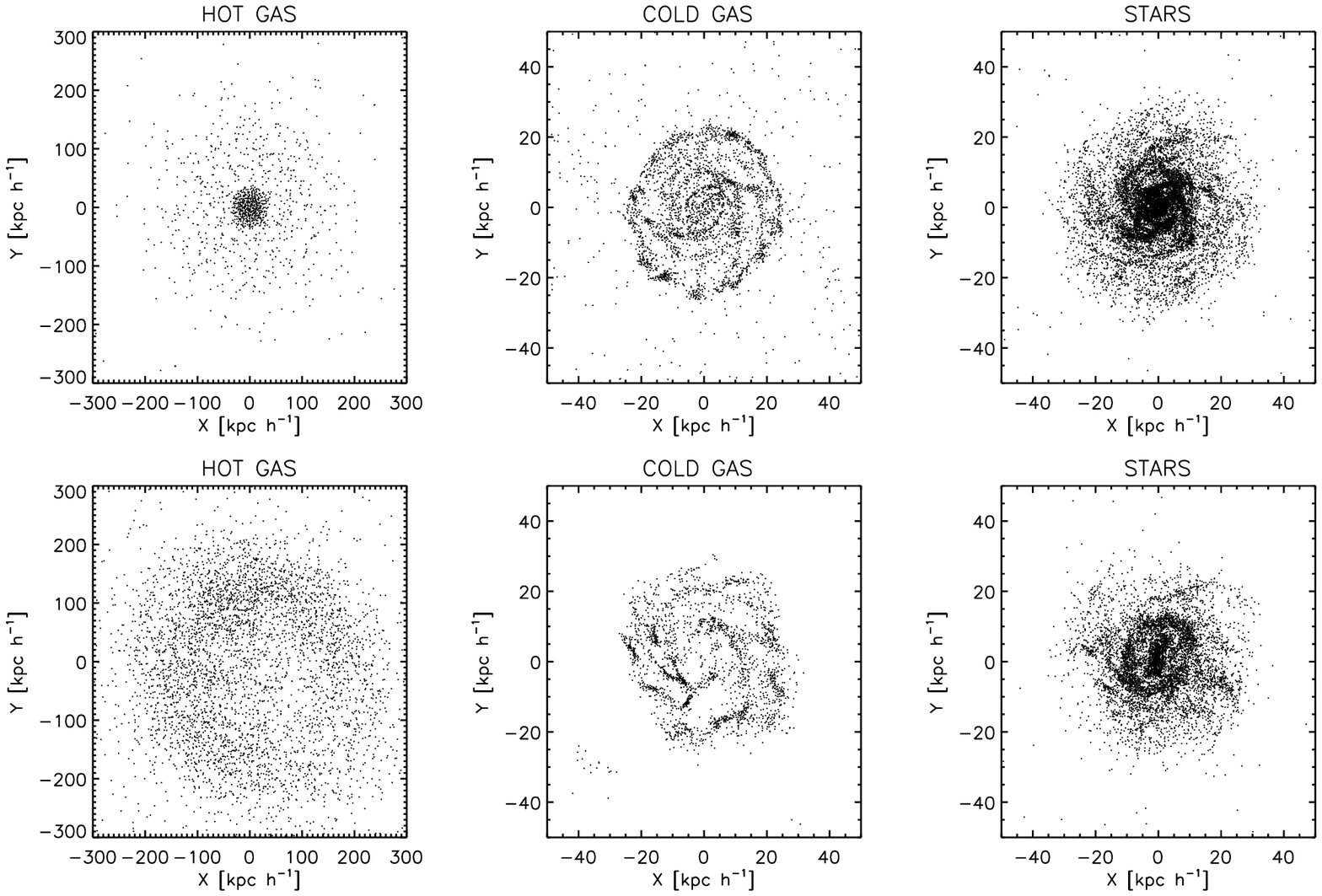}
\caption{Face-on projections of the discs in $10^{12}\, h^{-1}{\rm M}_\odot$
  systems after $1.2\, h^{-1}{\rm Gyr}$ of evolution.  We show the hot gas ($A
  \geq A_{\rm crit}$), the cold gas ($A < A_{\rm crit}$) and the stars
  separately, both for D12 (no energy feedback; upper panels) and for F12-0.5
  (SN feedback with $\epsilon_{\rm c}=\epsilon_{\rm h}=0.5$; lower panels).  }
\label{disk-feed}
\end{figure*}

In order to assess how the energy feedback model and the promotion scheme
work, we show in Fig.~\ref{promoted} the effects of promotion on individual
particles.  As explained earlier, each cold gas particle has been assigned a
reservoir in which we keep track of the energy it receives from nearby SN
explosions.  This reservoir energy is accumulated until the particle is
eventually promoted, dumping its reservoir into its internal energy.  In
Fig.~\ref{promoted}, we plot the properties of promoted particles in a
density-temperature plot for our test F12-0.5.  The arrows point from the
original location of each cold gas particle to its new position after
promotion, and the different panels correspond to particles promoted in
different time intervals.  It is clear that our feedback model is successful
in driving a gas flow from the cold, dense phase into the hot, diffuse medium
where the promoted particles share the properties of their local hot
environment. Note that the hot environment evolves with time, and
consequently, the new location of promoted particles on the
temperature-density plot changes accordingly. This flexibility arises because
we do not prescribe in advance the thermodynamic properties of the hot and
cold phases.

A further consequence of the reheating of cold gas and its promotion into the
hot phase is the generation of a significant outflow of material from the
system.  As an illustrative example, we show in Fig.~\ref{outflows} an edge-on
projection of the velocity field of promoted particles (arrows) in 
F12-0.5.
Different panels correspond to different times during the evolution of the
system: $t=0.3$ (upper left-hand panel), $0.4$ (upper right-hand panel), $0.75$ (lower
left-hand panel) and $t= 1.0\,h^{-1}{\rm Gyr}$ (lower right-hand panel). 
In each panel, the velocity field has been calculated using particles
promoted in the last $0.2\ h^{-1}$ Gyr.
The lengths of the arrows scale
with the magnitude of the local velocity, with the longest arrows
corresponding to $\sim 1000\,{\rm km\, s^{-1}}$.  The background colour
represents the projected metallicity for the total gas component.  From this
figure we can see that our promotion scheme establishes an effective mechanism
for transporting gas from the centre of the discs into the haloes. Once
particles are promoted, they are accelerated outwards, moving mainly in the
direction perpendicular to the disc plane.  As may be seen from this figure,
the outflow is not completely symmetric, since it is determined by the
local geometry of the gas and the stellar distribution.  Note that the
inclusion of energy feedback allows us to account for chemical enrichment of
the region outside the discs since the galactic outflows are able to transport
a significant fraction of the heavy elements (defined as all chemical elements
higher than hydrogen and helium) outwards.  We will come back to the impact of
SN energy feedback on the chemical properties of galaxies in
Section~\ref{metals}.

The exchange of mass between the cold and hot phases driven by our promotion
scheme strongly affects the evolution of the systems and their resulting mass
distributions.  In Fig.~\ref{disk-feed}, we show face-on projections of the
discs in D12 (no energy feedback; upper panels) and F12-0.5 (with energy
feedback; lower panels) at $t=1.2 \, h^{-1}{\rm Gyr}$.  We show the hot and
cold media defined in section 2.1 separately, as well as the stellar
component.  We can see from this figure that energy feedback helps to generate
a well-defined hot phase which is distributed out to $\sim 300\, h^{-1}{\rm
kpc}$ in this particular experiment.  In the case of no energy feedback (upper
panels) the hot phase is produced only by the collapse and virialization of
the system, is a small fraction of the gaseous mass, and remains bound within
$\sim 100\, h^{-1}{\rm kpc}$.  In this latter case, most of the gas has been
able to cool and condense, producing a large stellar clump in the centre.

\begin{figure}
\includegraphics[width=90mm]{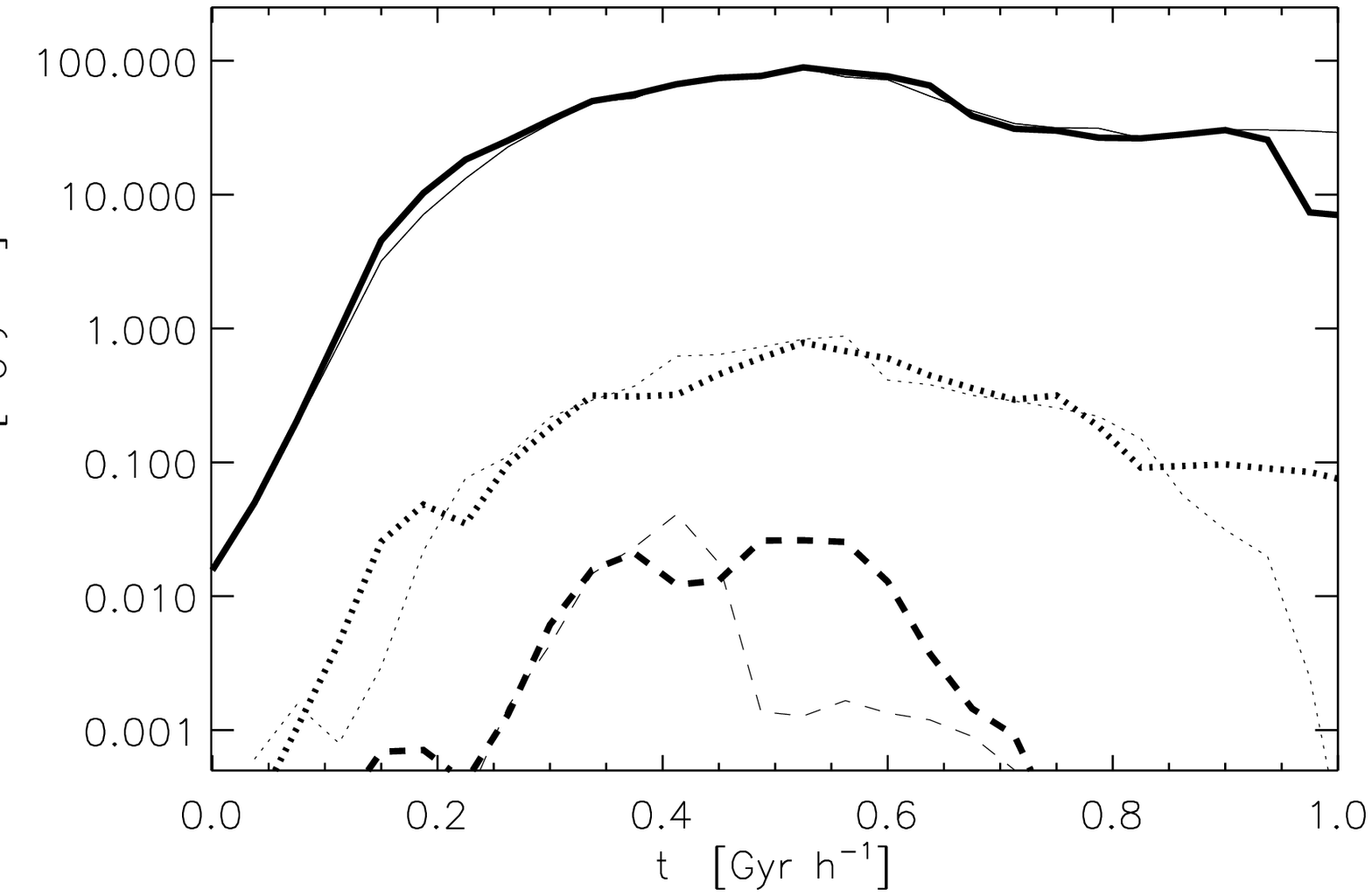}
\caption{Star formation rates for F12-0.5 (solid thin line), F12-0.5-H (solid
  thick line), F10-0.5 (dotted thin line), F10-0.5-H (dotted thick line),
  F9-0.5 (dashed thin line) and F9-0.5-H (dashed thick line), corresponding to
  idealized protogalactic collapses of differing total mass and numerical
  resolution.}
\label{resolution}
\end{figure}

\begin{figure}
\includegraphics[width=84mm]{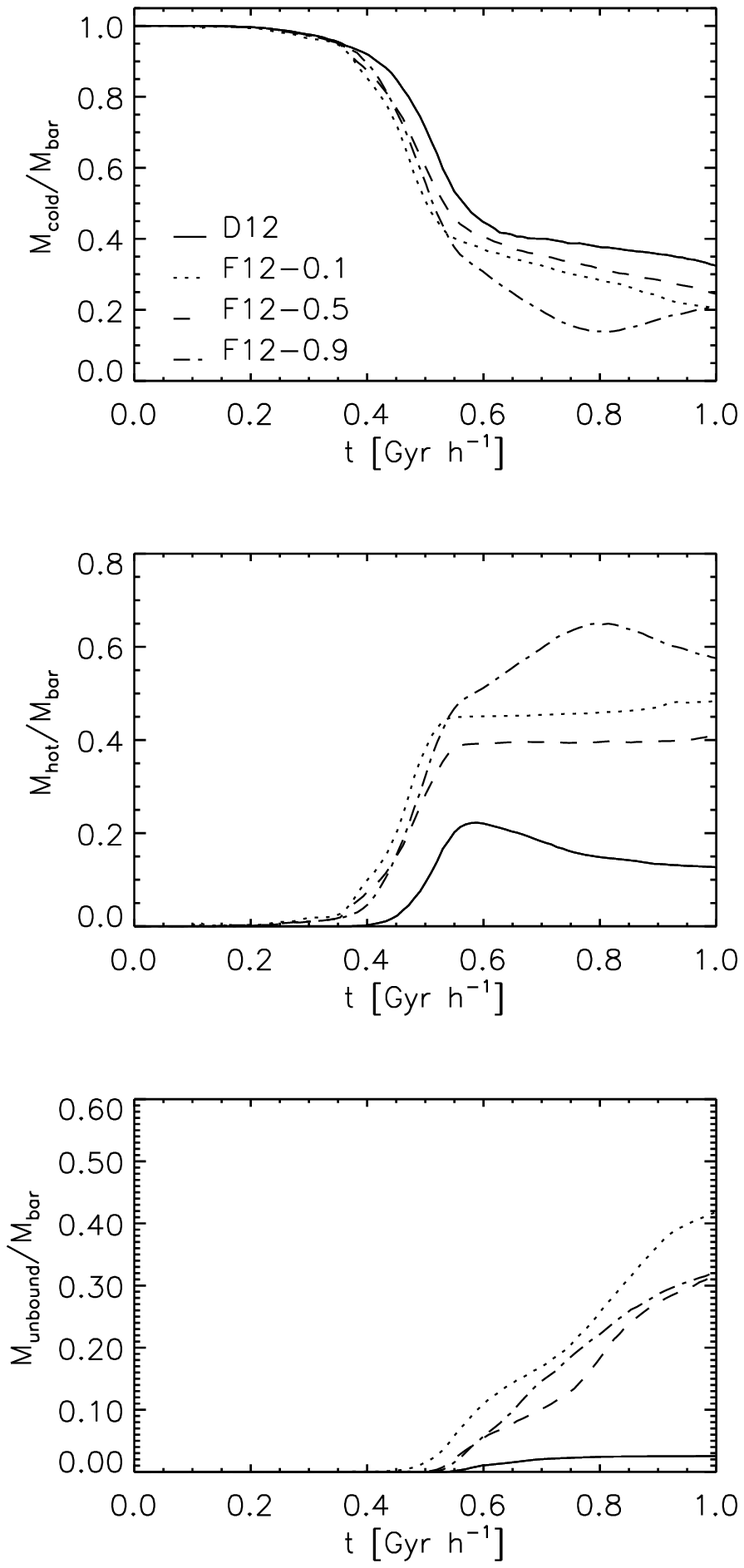}
\caption{Evolution of the mass fraction in cold gas ($A < A_{\rm crit}$), hot gas
  ($A \geq A_{\rm crit}$) and unbound gas, normalized to the total baryonic
  mass of the system for simulations D12 (no energy feedback) and F12-0.1,
  F12-0.5 and F12-0.9 (with energy feedback but with different $\epsilon_{\rm c}$
  values).}
\label{fig7-fcold}
\end{figure}

We have also tested how numerical resolution affects the evolution of systems
of different total mass. For this purpose we ran three additional simulations
increasing the initial number of gas and dark matter particles up to $40000$
in each mass component (see Table~\ref{simus-feedback}). These tests were
performed for systems with total masses of $10^{9}$, $10^{10}$ and
$10^{12}\,h^{-1}{\rm M}_\odot$ (F12-0.5-H, F10-0.5-H, F9-0.5-H, respectively),
using a feedback parameter of $\epsilon_{\rm c}=0.5$ in all cases.  In
Fig.~\ref{resolution}, we show the evolution of the star formation rate for
these models and compare them with their lower resolution counterparts.  From
this figure we can see that the results are reasonably well converged for the
higher halo masses.  For the smallest system, stochastic bursting behaviour is
observed so convergence is difficult to test.


\section{Analysis and Results}
\label{results}

In this section, we analyse the effects of our new feedback scheme on
simulations of galaxy formation.  In particular, we test the dependence of our
results on the adopted feedback parameters, and we investigate the impact of
the energy feedback on the resulting metal distributions. We also discuss the
dependence of the results on the total system mass.

\subsection{ Star formation and outflows}
\label{sfr-outflows}

In order to assess the response of galaxies to the adopted feedback
parameters, we compare simulations F12-0.1, F12-0.5 and F12-0.9 (see
Table~\ref{simus-feedback}) which only differ in the adopted value of
$\epsilon_{\rm c}$.  In Fig.~\ref{fig7-fcold}, we show the evolution of the cold ($A
< A_{\rm crit}$, upper panel) and hot ($A \geq A_{\rm crit}$, middle panel)
gaseous mass, in units of the total baryonic mass.  For comparison, we also
include the corresponding relations for simulation D12 which was run without
energy feedback.

We have also estimated the fraction of unbound gas (lower panel), defining it
here as the gas with a positive sum of kinetic and gravitational potential
energy.  As can be seen from this figure, the simulations with energy feedback
have larger fractions of hot and unbound gas compared with their counterpart
D12, indicating that the gas has been both heated and accelerated outwards.
Note, however, that the effect of varying the input parameter $\epsilon_{\rm c}$ is
not simple. This is a consequence of the non-trivial interplay between release
of SN energy, promotion from cold to hot gas, and radiative
cooling. All these processes influence the gas accretion and star formation
rates and the formation of the cold and hot phases.  As reference values we
give in Table~\ref{simus-feedback} the unbound gas fraction for all test
simulations after $1.0\,h^{-1}{\rm Gyr}$ of evolution.

\begin{figure*}
\includegraphics[width=154mm]{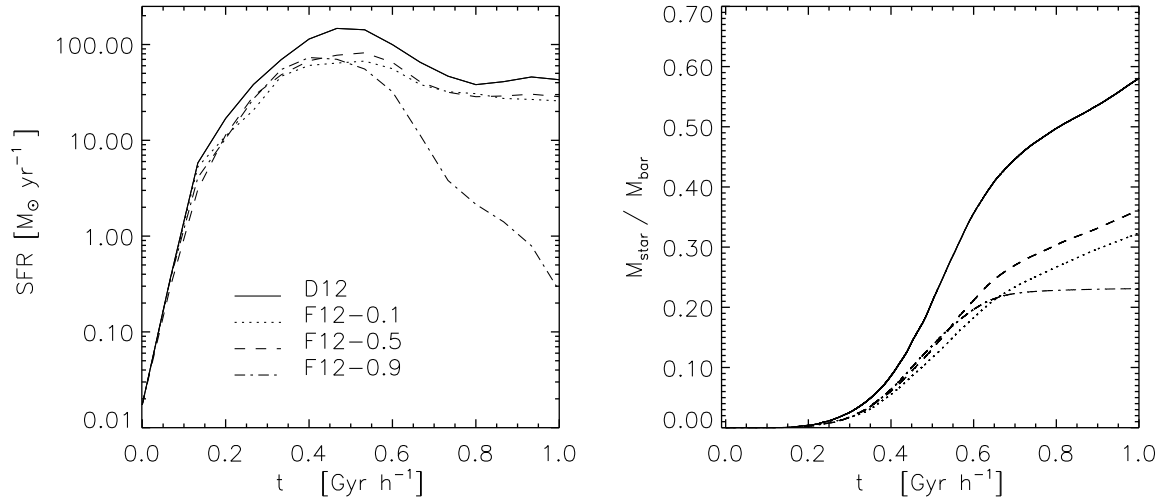}
\caption{Evolution of the star formation rate (left-hand panel) and
  integrated stellar mass fraction (right-hand panel) for D12 (no energy
  feedback), and F12-0.1, F12-0.5 and F12-0.9 (supernova energy
  feedback with different values of $\epsilon_{\rm c}$).}
\label{sfr-fcold}
\end{figure*}

As a consequence of the exchange of material between the phases, our scheme
can effectively regulate star-formation activity. This can be appreciated from
Fig.~\ref{sfr-fcold} where we show star formation rate and integrated stellar
mass fraction as a function of time for our simulation series.  Note that for
the initial conditions adopted here, particles initially located at the edge
of the system fall to the centre at $t \approx 0.55\, h^{-1}{\rm Gyr}$.
Compared to D12, supernova feedback reduces the final stellar mass by $\sim
45$, $40$ and $65$ per cent for F12-0.1, F12-0.5 and F12-0.9, respectively
(see also Table~\ref{simus-feedback}).  The larger $\epsilon_{\rm c}$, the more
energy injected into the cold phase, the sooner promotion becomes significant,
leading to a reduction in the amount of cold gas and the suppression of star
formation. On the other hand, as $\epsilon_{\rm c}$ increases, less energy is dumped
into the hot phase, making it easier both for hot gas to cool onto the disc
and for cold gas to be promoted. The interplay between these processes leads
to a non-monotonic dependence of the overall star-formation efficiency on
$\epsilon_{\rm c}$.

\begin{figure}
\includegraphics[width=84mm]{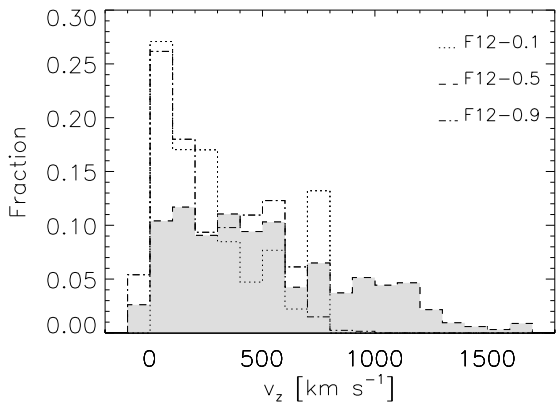}
\caption{Fraction of promoted gas mass as a function of $z$-velocity,
  defined here as positive if particles are flowing away from the centre and
  negative otherwise.  The plot shows results for our simulations of the
  $10^{12}\, h^{-1} {\rm M}_\odot$ system, run with different feedback
  parameters: F12-0.1 (dotted lines), F12-0.5 (dashed lines) and F12-0.9
  (dotted-dashed lines), at time $t = 0.8 \, h^{-1}{\rm Gyr}$ of the
  evolution.}
\label{hist_outflows}
\end{figure}

 Since in these tests promoted particles define the outflows,
we show in Fig.~\ref{hist_outflows} the mass fraction in promoted gas
as a function of $z$-velocity $v_{\rm z}$, defined as positive if particles
are moving away from the centre and negative otherwise. Results are given for
F12-0.1, F12-0.5 and F12-0.9 after $0.8\, h^{-1}{\rm Gyr}$ of evolution (just
after the maximum in star formation activity).  
For our test without SN
energy feedback (D12), we calculated that $89$ per cent of the gaseous mass
has $|v_{\rm z}| < 100\, {\rm km\, s^{-1}}$, with an upper limit of $|v_{\rm
z}| \approx 500 {\rm \ km\ s^{-1}}$.  In contrast, as may be seen in
Fig.~\ref{hist_outflows}, in our simulations with feedback the fraction of
{\it promoted} gas with $v_{\rm z} > 500 {\rm \ km\ s^{-1}}$ varies from 22 to
64 per cent.  Clearly our scheme can produce outflows with relatively high
velocities; the upper limits have increased to $v_{\rm z} > 1000\, {\rm km\,
s^{-1}}$.  We can also infer from this figure that the fraction of promoted
gas which flows back into the system (indicated by negative velocities) is
small.
Observationally, the
velocities associated with galactic outflows are found to vary from a few
hundred to thousands of kilometres per second (e.g. Lehnert \& Heckman 1995,
1996).  Here, we do not intend to make a detailed comparison with observations
but to investigate the typical $z$-velocities that the gas components
associated with outflows may reach in our model.  
Our estimates show that the simulated outflows can reach $z-$velocities in
the observed range. Note, however, that the correspondence between the observed
line shifts and the velocities we measure in our simulations is not straightforward
because of the variety of hydrodynamical, ionization and radiative transfer
processes occuring in real systems.

\subsection{Metal distribution}
\label{metals}

The redistribution of the gas owing to the injection of energy by SNe
also affects the chemical abundances of the cold and hot gas and those of the
stars. It is thought that SN feedback could be the mechanism responsible for
the transport of heavy elements into the intergalactic medium. In order to
study this process in detail, it would be necessary to simulate galaxies in
their correct cosmological setting, which goes beyond the scope of this paper.
As a first step, we can however investigate the potential effects of our
feedback model on the metal distribution using our idealized simulations of
protogalactic collapses of total mass $10^{12}\, h^{-1}{\rm M}_\odot$ (see
Table~\ref{simus-feedback}).

\begin{figure*}
\includegraphics[width=154mm]{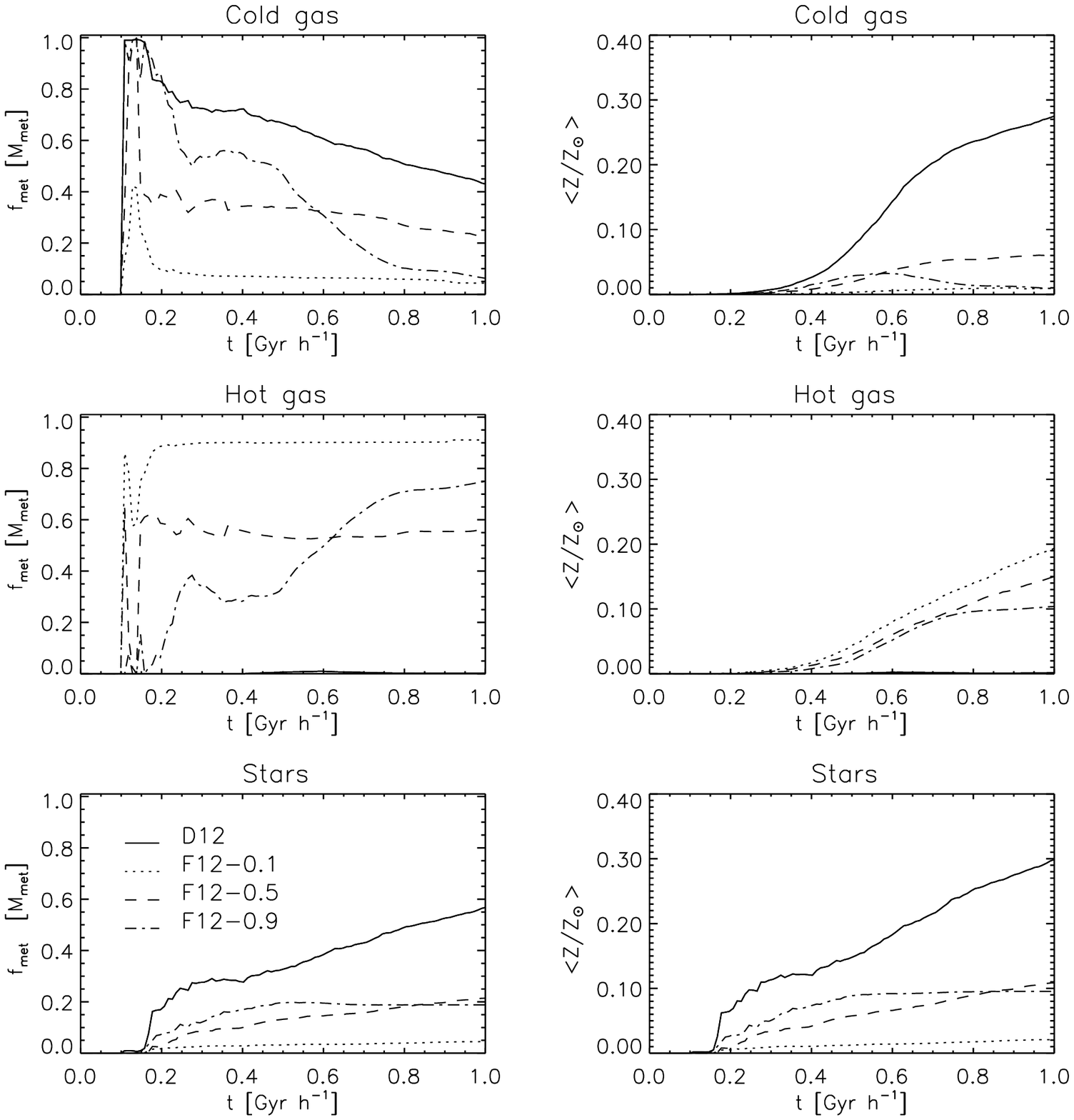}
\caption{Evolution of the fraction of metals 
  in the cold gas ($A < A_{\rm crit}$), hot gas ($A \geq A_{\rm crit}$) and in
  the stars (left-hand panels), for our $10^{12}\, h^{-1} {\rm M}_\odot$ mass
  system. We show results for simulation D12 without energy feedback, and for
  simulations F12-0.1, F12-0.5 and F12-0.9 corresponding to different feedback
  parameters.  We also show the evolution of the mean metallicity for the same
  runs (right-hand  panels).}
\label{metals-fases}
\end{figure*}

In Fig.~\ref{metals-fases} (left-hand panels), we show the evolution of the
fraction of metals locked into cold gas ($A < A_{\rm crit}$), hot gas ($A \geq
A_{\rm crit}$) and stars, normalized to the final total mass in heavy
elements, for D12, F12-0.1, F12-0.5, and F12-0.9.  In the right-hand panels of
Fig.~\ref{metals-fases}, we also display the mean metallicity $Z$, in units of
solar metallicity, for these same simulations.  Note that the total mass in
metals in the system is a function of time owing to ongoing star formation.
From this figure we can see that if energy feedback is not included, the metal
content remains completely locked into the stellar and cold gas
components. This is expected, since  there is no efficient
mechanism to transport metals away from the star-forming ISM.  This situation is
radically changed by the powerful winds generated by the SN feedback scheme. These
outflows redistribute
the metals, increasing the amount of enriched hot gas at the expense of the
cold gas and stars.  Note that even in F12-0.9, for which only $10$ per cent
of the metals are directly injected into the hot phase, the dynamical and
thermal evolution of the system differs markedly from the case without energy
feedback, showing efficient enrichment of the hot gas.  This is driven by
promoted material which was enriched while cold.  In the case of F12-0.1, the
cold gas and the stars end up with a negligible fraction of the metals.  These
trends are also reflected in the mean metallicities of the gaseous and stellar
components.  Fine-tuning $\epsilon_{\rm c}$ and $\epsilon_{\rm r}$ in a simulation with
self-consistent cosmological initial conditions may allow one to reproduce the
observed metallicity trends for hot gas, cold gas and stars.  We will attempt
this for a Milky Way look-alike in a forthcoming paper.

\subsection{Dependence on virial mass}
\label{mvirial}

From a theoretical point of view, SN feedback is expected to play a
crucial role in regulating the star formation activity in galaxies and in
determining the level of enrichment of different environments.  It is also
expected that the impact of SN feedback should depend strongly on
virial mass, where smaller mass systems should be more strongly affected,
their shallower gravitational potential wells making it easier for the
supernovae to drive outflows (Larson 1976; White \& Rees 1978; Dekel \& Silk
1986).  This concept has been widely applied in semi-analytic models of galaxy
formation to explain several basic observational properties of the galaxy
population such as the faint-end slope of the luminosity function (e.g.  White \&
Frenk 1991).  In previous sections, we have demonstrated that our energy
feedback model is able to produce gas outflows which  lead to different
levels of enrichment, depending on the adopted $\epsilon_{\rm c}$ value.  In this
section, we discuss the effects of SN energy feedback as a function of the
mass of the systems considered.

\begin{figure*}
\includegraphics[width=154mm]{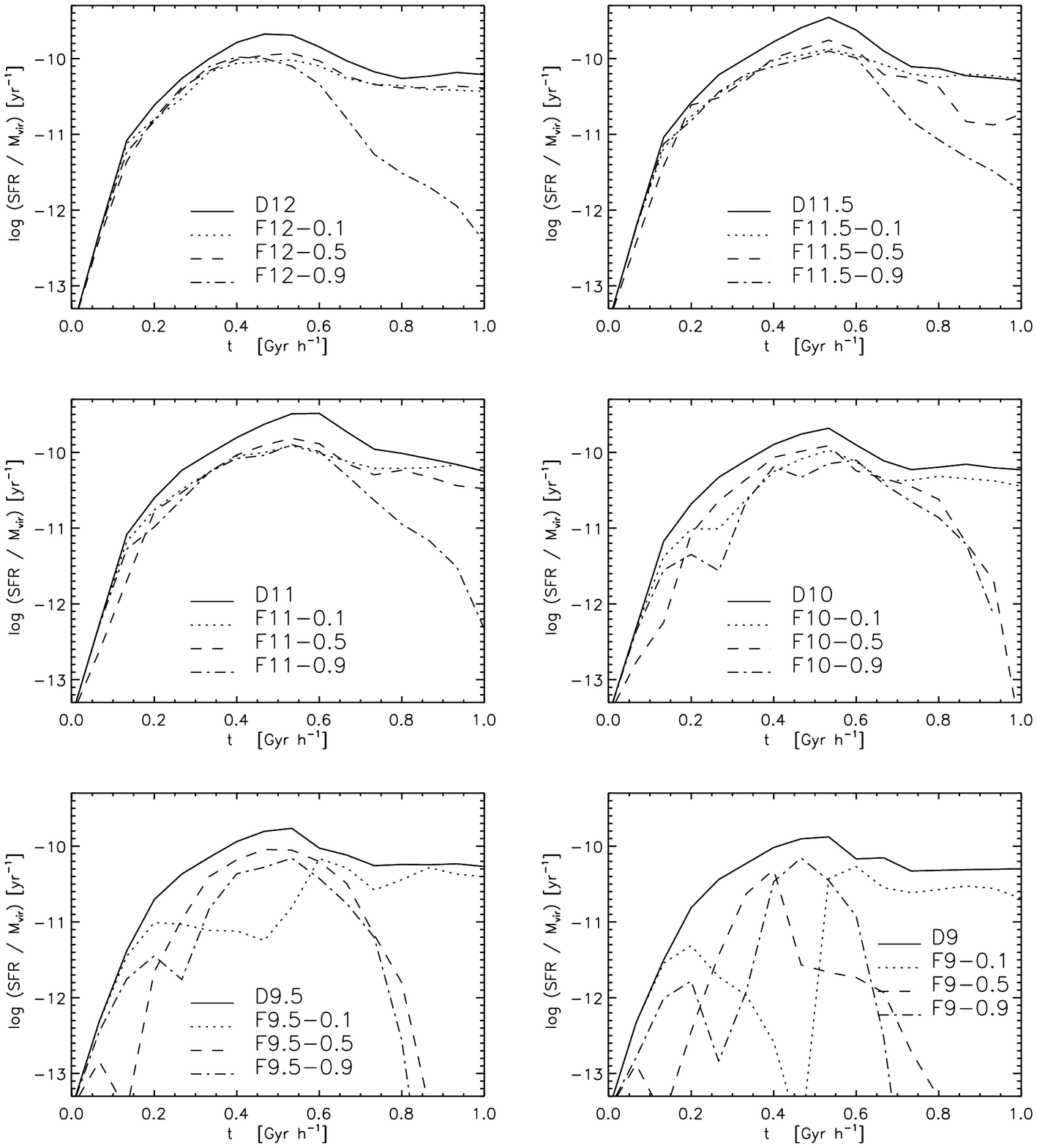}
\caption{Star formation rates normalized to virial mass for simulations with total masses of $10^{9}$,
$10^{9.5}$, $10^{10}$, 
 $10^{11}$, $10^{11.5}$, and $10^{12}\,h^{-1} {\rm M}_\odot$.  The plots give results both
for simulations without SN feedback (solid lines), and for simulations
with SN feedback with $\epsilon_{\rm c}=0.1$ (dotted lines), $\epsilon_{\rm c}=0.5$
(dashed lines) and $\epsilon_{\rm c}=0.9$ (dashed-dotted lines).   }
\label{sfr_mvir}
\end{figure*}

We again start from our idealized spherical initial conditions for
protogalactic collapse.  These initial conditions are simple enough to
highlight the effects of feedback without being distracted by additional
processes such as mergers and infall, which complicate the picture in fully
hierarchical scenarios for galaxy formation. We analyse results from
simulations that cover the total mass range $10^{9}-10^{12}\,h^{-1}{\rm
M}_\odot$, as well as different choices of the feedback parameters (see
Table~\ref{simus-feedback}).  For comparison, we have also analysed simulations
without feedback.

Fig.~\ref{sfr_mvir} compares star formation histories for systems of
differing total mass, in each case for simulations without SN energy feedback
(solid lines) and with the feedback model enabled, using $\epsilon_{\rm c}=0.1$
(dotted lines), $\epsilon_{\rm c}=0.5$ (dashed lines) and $\epsilon_{\rm c}=0.9$
(dashed-dotted lines).   Clearly, SN feedback has a strong impact on these
systems regardless of their total mass.  For larger masses, the primary effect
of SN feedback is a decrease in star formation activity. As one moves to lower
mass systems, SN feedback modifies the star formation rates not only by
decreasing the overall level of activity, but also by changing their
character, producing a series of starbursts.  From Fig.~\ref{sfr_mvir}, we can
also infer that increasing the value of $\epsilon_{\rm c}$ generally leads to
stronger feedback effects at all times.  In particular, larger $\epsilon_{\rm c}$
values are able to trigger several starburst episodes and to produce a
significant decrease in the star formation activity at later times.
 Note that, specially for the smallest galaxies, star formation can
stop  completely after $~\sim 1\ h^{-1}$ Gyr because the systems can lose
their gas  reservoirs. However, if more realistic initial conditions
were considered,
gas accretion from the intergalactic medium could contribute to fuel
new star formation activity at later times.

The results obtained for our smallest virial masses, in which the star formation histories
show  episodic, bursty behaviours, are consistent with 
observational studies of dwarf galaxies (e.g. Kauffmann et
al. 2003; Tolstoy et al. 2003). 
Numerical simulations assuming different types
of initial conditions 
have also found such bursty behaviour for the star formation
activity  in small mass systems (e.g. Steinmetz \& Navarro 1999; Carraro et al. 2001; Mayer
et al. 2001; Pearce et al. 2001; 
Chiosi \& Carraro 2002; Springel \& Hernquist 2003). 
Some of these papers assume CDM universes in which star
formation is mainly triggered by mergers, while others 
assume galaxies to form by monolithic collapse of baryons inside virialized
dark matter haloes.
Although most of them have considered the effects of SN feedback,  none has succeeded 
in  establishing a self-regulated star formation activity without
introducing scale-dependent paramaters.

\begin{figure}
\includegraphics[width=84mm]{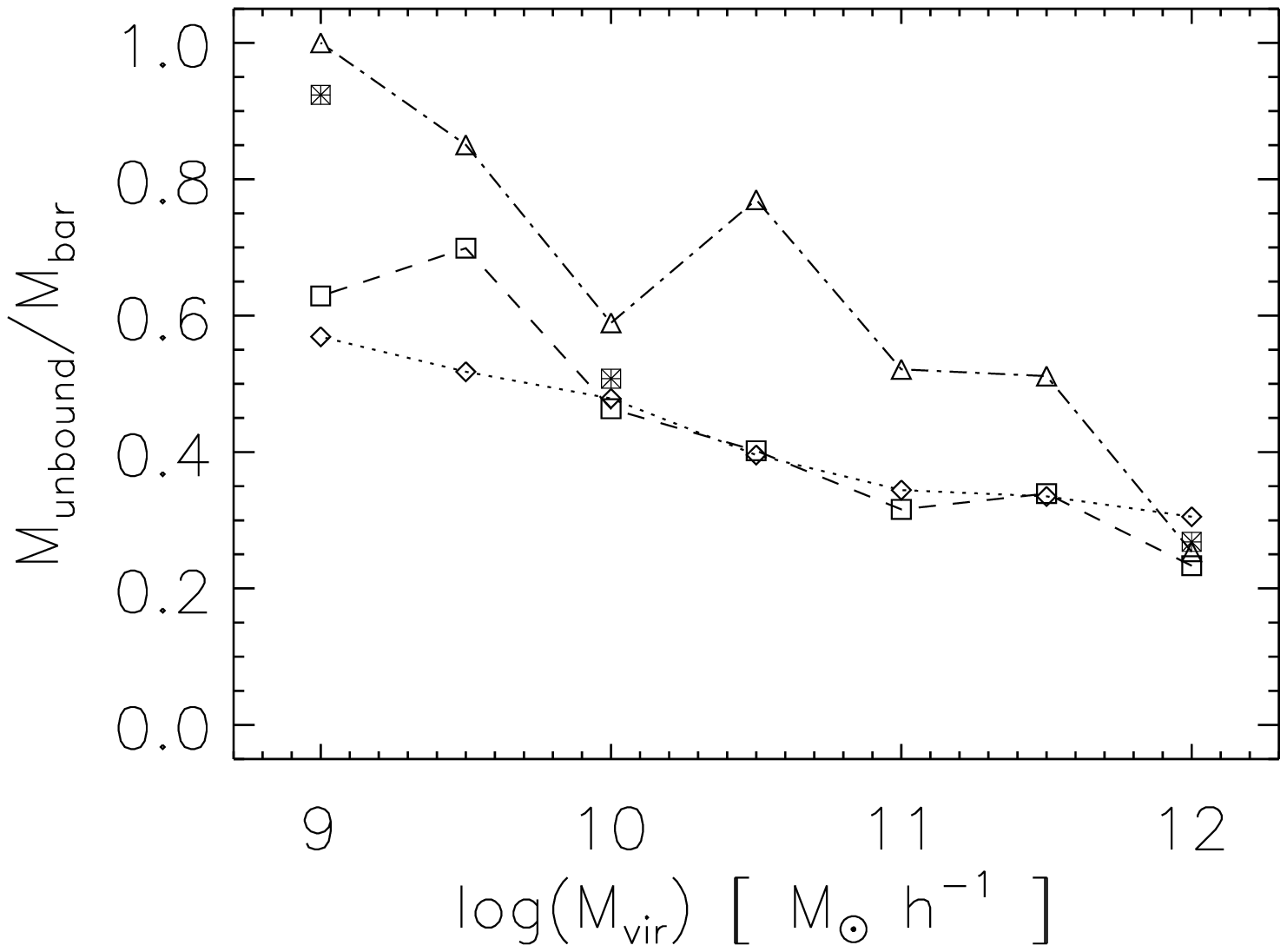}
\caption{Fraction of unbound gas as a function of total baryonic mass for simulations
  with $\epsilon_{\rm c}=0.1$ (dotted line), $\epsilon_{\rm c}=0.5$ (dashed line) and
  $\epsilon_{\rm c}=0.9$ (dashed-dotted line) after $0.85\,h^{-1} {\rm Gyr}$ of 
  evolution.  We also show the results for our higher resolution simulations
  with $\epsilon_{\rm c}=0.5$ (squared asterisks).}
\label{funbound_mvir}
\end{figure}

In Fig.~\ref{funbound_mvir}, we show the fraction of unbound gas as a
function of virial mass after $0.85\,h^{-1}{\rm Gyr}$ of evolution, for
simulations with $\epsilon_{\rm c}=0.1$ (dotted line), $\epsilon_{\rm c}=0.5$ (dashed line)
and $\epsilon_{\rm c}=0.9$ (dashed-dotted line).  We also show the results for our higher
resolution simulations with $\epsilon_{\rm c}=0.5$ (squared asterisks).  As
discussed earlier, the unbound gas fraction can be viewed as a proxy for
feedback strength since it quantifies the fraction of baryons that are swept
away from the system as a result of a feedback-driven outflows.  For a given
total mass, as the value of $\epsilon_{\rm c}$ is increased, the fraction of unbound
gas increases.  If no feedback is included, the fraction of unbound gas is
negligible.  As expected, for a given value of $\epsilon_{\rm c}$, SN feedback has
stronger effects as we go to smaller systems.  The smaller the system, the
larger the fraction of unbound gaseous mass and the smaller the final stellar
mass fraction (see Table~\ref{simus-feedback}).  These trends with total mass
are very encouraging, suggesting that our feedback model may reproduce the
shape of the faint end of the galaxy luminosity function in hierarchical
clustering scenarios (e.g. Trentham \& Tully 2002). This has been an elusive
goal in high-resolution simulations of galaxy formation thus far.  The
self-consistent regulation of the star formation activity together with the
natural production of vigorous outflows can be viewed as an important
achievement of our new feedback model.

\begin{figure}
\includegraphics[width=84mm]{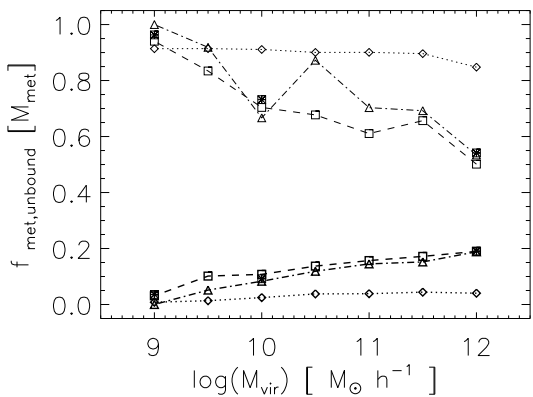}
\caption{Fraction of metals locked into unbound gas (thin lines) and into
  stars (thick lines) as a function of total mass for simulations with
  $\epsilon_{\rm c}=0.1$ (dotted lines), $\epsilon_{\rm c}=0.5$ (dashed lines) and
  $\epsilon_{\rm c}=0.9$ (dashed-dotted lines) after $0.85\,h^{-1}{\rm Gyr}$ of evolution.
  We also show results for our higher resolution simulations with
  $\epsilon_{\rm c}=0.5$ (squared asterisks).}
\label{fhotcoldmvir}
\end{figure}

Finally, in Fig.~\ref{fhotcoldmvir} we show the fraction of metals locked
into the unbound gas (thin lines) and into the stellar component (thick
lines), as a function of total mass after $0.85\, h^{-1}{\rm Gyr}$.  We
show the results for simulations with $\epsilon_{\rm c}=0.1$ (dotted line),
$\epsilon_{\rm c}=0.5$ (dashed line) and $\epsilon_{\rm c}=0.9$ (dashed-dotted line) and for our
higher resolution tests with $\epsilon_{\rm c}=0.5$ (squared asterisks).  From this
figure we can see that the unbound gas is highly enriched, even in the largest
mass systems.  The anticorrelation in this figure indicates that the
fraction of metals locked into the unbound gas increases with decreasing
total mass.  In contrast, the fraction of metals locked into stars increases
slightly with the virial mass. In this case, note that the fraction of metals
locked into the stellar component is less than $20$ per cent which  is perhaps
suggestive of the need for a top-heavy IMF (Nagashima et al. 2005).  This
could help to reach higher abundances for the stellar populations, in better
agreement with observation (e.g. Gallazzi et al. 2005).  Also note
that in the simulations with $\epsilon_{\rm c}=0.1$, most of the metal content of
the system is located in the unbound gas.
 
The above results suggest that smaller systems would contribute to the
enrichment of the intergalactic medium with larger fractions of their metal
production.  However, as large systems also produce important fractions of
unbound and enriched gas, the amount of metals they release into the
intergalactic medium could be very significant or even dominant.  This
behaviour may explain recent observational findings on the stellar
mass-metallicity relation (Tremonti et al. 2004) where fewer metals were
detected in massive galaxies than expected for closed box models,
demonstrating rather directly that galaxies of all masses may have lost heavy
elements to their surroundings.

The dependence of the impact of SN energy input on galaxy escape
velocity may be able to explain the observed mass-metallicity
relation in the context of a  $\Lambda$CDM universe (e.g. De Lucia, Kauffmann \& White 2004).
Using  the chemically enhanced GADGET-2 of Paper I (but without our new feedback
scheme), Tissera, De Rossi \& Scannapieco (2005)
found that in the absence of strong  winds a
mass-metallicity relation is produced which is less steep
at low masses than that observed (e.g. Tremonti et al. 2004). In  future work,
we will explore the mass-metallicity
relation of galaxies in full cosmological simulations using our new SN energy feedback
model.

\section{Star formation efficiency and the Kennicutt law}
\label{kennicutt}

Our decoupling, star formation and feedback schemes contain a number of
numerical and physical parameters for which values must be chosen. We
discussed the phase separation parameters $\alpha$, $T_*$ and $\rho_*$ in
sections~\ref{decoupling} and ~\ref{feedback}, 
arguing that the behaviour of our algorithms is not
sensitive to their precise values; $\alpha$ is chosen to achieve effective
decoupling of gas at very different entropies while the values of density and
temperature which define cold gas for the purpose of our feedback algorithms are
based on the shape of the radiative cooling curve and the properties of
observed star-forming regions.  In addition, values must be chosen for the
efficiencies of star formation and of energy deposition in hot and cold gas,
for the chemical yields from stellar evolution, and for parameters which
regulate the distribution of metals between the gas components. In previous
sections we explored the effects of varying the efficiencies with which energy
is deposited in hot and cold gas and we tied the distribution of metals to
that of energy. In this section we determine appropriate values for the star
formation efficiency.

Observationally, it is found that star formation rate per unit area is
remarkably tightly correlated with total gas surface density (Kennicutt
1998). This purely empirical relation holds as a function of radius within
galaxies as well as among galaxies of very different mass, type and
redshift. It is a considerable challenge for galaxy formation models such as
our own to reproduce this 'Kennicutt Law', since we have freedom only to
adjust the overall efficiency of star-formation; the slope and the
surprisingly small scatter of the relation must emerge 'naturally' from the
self-regulation of star formation within the numerical model.  Below we show
that our model can indeed reproduce the observed behaviour both for the
simplified collapse model used in earlier sections and for an equilibrium
galaxy model with structure similar to that of the Milky Way.

\begin{figure}
\includegraphics[width=84mm]{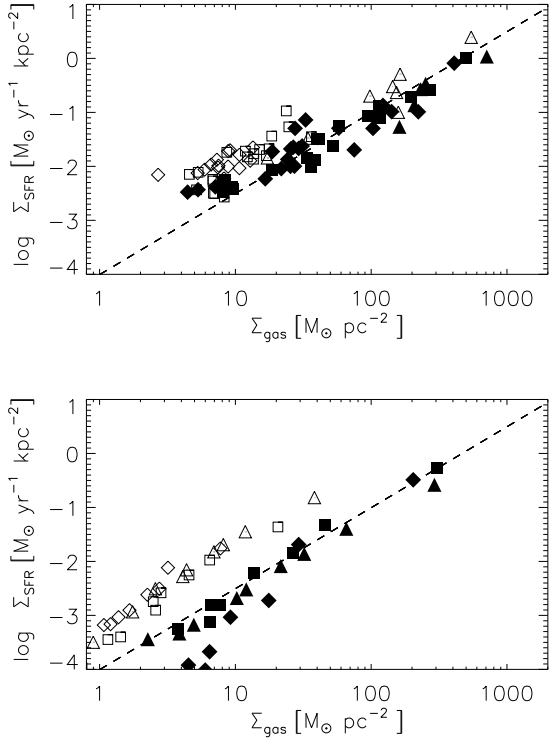}
\caption{The upper panel shows projected star formation rate density as a
  function of gas surface density for two simulations of the collapse of a
  $10^{12}\, h^{-1}{\rm M}_\odot$ mass system. Both surface densities are
  averaged over a series of annuli in face-on projections of the
  simulations. Open symbols refer to the model F12-0.5 discussed above at
  times 0.5 (triangles), 1.0 (squares) and 1.5 Gyr (diamonds), while
  corresponding filled symbols refer to the same times and to an otherwise
  identical model with $c=0.01$. The lower panel shows results for two
  simulations from equilibrium initial conditions representing a spiral galaxy
  with structure similar to that of the Milky Way. The symbols refer to the same
  times and  values of $c$ as in the upper panel. The dashed line in both
  panels represents the relation of Kennicutt (1998).}
\label{kennicutt_law}
\end{figure}

Our star formation algorithm assumes a star formation rate per unit
volume equal to
\begin{equation}
\dot\rho_\star = c\,\frac{\rho}{\tau_{\rm dyn}} , 
\end{equation}
where $\rho_\star$ is the density of the new born stars, $\rho$ is the gas
density, $\tau_{\rm dyn}$ is the dynamical time of the gas, defined simply as
$\tau_{\rm dyn} = 1/(4\pi G\rho)^{1/2}$, and $c$ is a star formation efficiency
parameter (see Paper I for details).  In the numerical experiments analysed
above, we used $c=0.1$.  In this Section, we investigate the value required to
reproduce the Kennicutt Law by comparing our previous collapse model F12-0.5,
with a virial mass of $10^{12}\ h^{-1}$M$_\odot$, to an otherwise identical
model where we lowered the star formation efficiency to $c=0.01$. In the upper
panel of Fig.~\ref{kennicutt_law} we show the resulting projected surface star
formation rates and surface gas densities at three times, 0.5, 1.0 and 1.5
Gyr, averaging over broad annuli out to a radius of $30\ h^{-1}$ kpc. The
observed Kennicutt relation is indicated on this plot by a dashed straight
line.  Clearly, for $c=0.1$ (the open symbols) star formation is much more
efficient in the simulation than is observed; disc gas is consumed rapidly,
and there is considerable scatter about the mean relation. Reducing $c$ by an
order of magnitude reduces the star formation rates at given surface density
by a similar factor, resulting in a much broader distribution of gas surface
densities and an excellent fit to the zero-point slope and scatter of the
observed relation. In addition, it is clear that as our idealized system
builds up its disc it evolves along the Kennicutt relation. It is interesting
that although star formation activity is reduced substantially for $c=0.01$,
the SN feedback model is still able to produce a strong galactic wind. After
$1.5$ Gyr of evolution, we find that $\sim10$ per cent of all
baryons are both unbound and enriched with heavy elements.

The collapse initial conditions used here and above are highly idealized. In
order to assess whether the apparent agreement with observation extends to
quiescent star formation in an equilibrium system, we analyze two further
simulations from initial conditions representing a spiral galaxy similar in
structure to the Milky Way. The initial system is based on that studied in
Springel, Di Matteo \& Hernquist (2005). It consists of a dark matter halo, a stellar
bulge, a diffuse hot-gas halo, and a disc with $10$ per cent of its mass in
gas and the rest in stars.  The dark matter distribution is modelled as a
spherical NFW distribution with a circular velocity at $r_{200} = 160
h^{-1}{\rm kpc}$ of $161$ km s$^{-1}$ and a concentration parameter of $9$
(Navarro, Frenk \& White 1996, 1997).  The disc has a radial scale-length of
$2.5\, h^{-1}{\rm kpc}$, a vertical scale-height of $0.5\, h^{-1}{\rm kpc}$,
and a Toomre stability parameter of $Q>1$ for the stars.  The spherical bulge is
taken to have an exponential profile with scale-length $0.2$ times that
of the disc. The density of the hot gas follows that of the dark matter. The
bulge, disc, hot gas halo and dark matter halo are taken to have initial
masses of 0.13, 0.39, 0.059  and $9.0\times10^{11}\, h^{-1}{\rm M}_\odot$
respectively.  Initially dark matter, gas and stars are represented by
simulation particles with mass 300, 2.0 and $15\times 10^5\,h^{-1}{\rm
M}_\odot$ respectively.  We have used gravitational softenings 
of $0.16$,  $0.32$ and $0.2\ h^{-1}$ kpc
for gas, dark matter and star particles, respectively.
In these simulations, we assume that bulge and disc stars are old
(i.e.  they do not contribute to metal and energy ejection). Metals 
and SN energy are exclusively produced by the new stars 
which can be formed from the $10$ per cent of the initial gas component.
The system is set up in equilibrium and we followed
its evolution using our full phase separation, star formation and feedback
models with feedback parameters $\epsilon_{\rm r}=0$ and $\epsilon_{\rm c}=0.5$ and with
$c=0.1$ and $c=0.01$. The star formation physics thus corresponds to the two
versions of F12-0.5 compared above.

The lower panel of Fig.~\ref{kennicutt_law} shows the Kennicutt relations we find
for these simulations at three different times. Despite the difference in
initial conditions, the results resemble those found already for F12-0.5. For
$c=0.1$ the disc gas is used up quite rapidly (only 30, 17 and 11 per cent of
the initial cold disc gas remains in the disc at the three times shown) and
the star formation rates lie well above the empirical Kennicutt relation. With
the lower star formation efficiency the observational relation is matched
quite well and gas consumption times are lengthened (81, 67 and 54 per cent of
the initial cold disc gas now remains in the disc at the three times
shown). It is remarkable that even with the relatively modest star formation
rates now found in this model, a significant galactic wind is generated. At
the final time ($t = 1.5$ Gyr), 51 per cent of all metals formed are in
the hot halo gas and 99 per cent of this gas is formally unbound and will
likely leave the galaxy.  Dividing this unbound mass by the age of the system
leads to an estimated mass-loss rate in the wind of $6\ {\rm M_\odot\ yr^{-1}}$
which can be compared with the mean star formation rate of $2\  {\rm M_\odot\
yr^{-1}}$ averaged over the duration of the simulation. The ratio of these two
rates is gratifyingly close to observational estimates for real wind-blowing
galaxies (Martin 1999, 2004). 
The correct representation of the Kennicutt law found for 
$c=0.01$ suggests this star formation efficiency as a good choice.
The tests of this section confirm that our new
SN-feedback model can match the observed phenomenology of star formation both
during starburst episodes and during phases of quiescent star formation.

\section{Conclusions}\label{conclusions}

We have developed a new model for energy feedback by SNII and SNIa within the
cosmological TreeSPH code {\small GADGET-2}, complementing the implementation
of chemical enrichment and metal-dependent cooling described in Paper~I.  The
model is tied to a multiphase treatment of the gas components in the ISM,
designed to improve the description of the hot, diffuse material in the
context of the SPH technique.  One of the major advantages of our model is
that no scale-dependent parameters need to be introduced.
This makes it especially well suited for simulations of cosmological
structure formation where systems of widely different mass form naturally.

We have used a number of idealized simulations of the formation of disc
galaxies to study the performance of our numerical techniques.  We found that
our multiphase scheme leads to an improved description of the diffuse, hot gas
which can reach densities up to an order of magnitude higher than those found
with a standard SPH implementation of {\small GADGET-2}.  The multiphase
scheme is insensitive to its single free parameter $\alpha$ when the value is
varied within a physically motivated range.

Our SNe feedback scheme efficiently regulates star formation by
reheating cold gas and generating winds.  The overall impact of SN feedback
depends on total mass: in large systems the inclusion of feedback reduces star
formation by a factor of a few compared to simulations where it is absent.
Smaller systems are more strongly affected, with star formation reduced by
more than an order of magnitude and occurring in stochastic starburst
episodes.  This finding  is consistent with observational studies which detect bursty
behaviour in the star formation activity of dwarf galaxies (e.g. Kauffmann et
al. 2003; Tolstoy et al. 2003).  
In the simulations presented here, 
this behaviour has been achieved by applying our self-consistent 
multiphase gas and SN energy
feedback schemes without  introducing scale-dependent parameters.
Furthermore,
our SN feedback model can
reproduce the Kennicutt relation between the surface densities of gas and of
star formation both in collapsing and in quiescent systems.

For all the masses we considered, the star formation histories of our
idealized protogalactic collapse models are sensitive to the fraction
$\epsilon_{\rm c}$ of energy and metals which is dumped into the cold phase. In
general, the larger the value of $\epsilon_{\rm c}$, the more star formation is
suppressed, but the effects of varying this parameter are complex because of
the non-linear coupling between star formation, SN energy release, and
radiative cooling.

Our SN feedback model results in strong winds from star-forming
galaxies.  The outflows are generally perpendicular to the disc plane and can
reach velocities of up to $1000\, {\rm km\, s^{-1}}$.  We detect a clear
anti-correlation between the unbound gas fraction and the total mass,
indicating that the strength of the outflows is sensitive to the depth of the
gravitational potential well, as expected from theoretical considerations.
For a given total mass, we find that the unbound gas fraction increases with
the assumed value of the feedback parameter $\epsilon_{\rm c}$. The mass-loss rates
we find typically exceed the star formation rates by a factor of a few, as
observed in starbursting galaxies.

These winds are efficient in enriching the outer regions of our simulated
galaxies and would plausibly contaminate the intergalactic medium.  We found
that the unbound gas component, regardless of total mass, ends up containing
more than $60$ per cent of the metals produced in our protogalactic
collapses. For the smallest galaxies, this percentage goes up to $80$ per cent
or more, independent of the assumed value of $\epsilon_{\rm c}$.  Galaxies with a
wide range of masses may thus contribute to the enrichment of the
intergalactic medium. This appears required to explain the observation that
the bulk of the heavy elements in galaxy clusters resides in the intracluster
medium rather than in the galaxies themselves. It will be extremely
interesting to compute detailed predictions for this enrichment using
cosmological simulations based on our feedback scheme.

\section*{Acknowledgements}

 We thank the anonymous referee for a careful and constructive
report.
This work was partially supported by the European Union's ALFA-II programme,
through LENAC, the Latin American European Network for Astrophysics and
Cosmology. Simulations were run on Ingeld and HOPE PC-clusters at the Institute
for Astronomy and Space Physics. We acknowledge support from Consejo Nacional
de Investigaciones Cient\'{\i}ficas y T\'ecnicas and Fundaci\'on Antorchas.
C. Scannapieco thanks the Alexander von Humboldt Foundation, the Federal Ministry of
Education and Research and the Programme for Investment in the Future (ZIP) of
the German Government for partial support.  The authours thank the Aspen Center
for Physics where part of the discussions of this work took place.

\end{document}